\documentclass[pre,twocolumn,showpacs,superscriptaddress,aps,floatfix]{revtex4-2}

\usepackage{amsfonts}
\usepackage{amsmath}
\usepackage{amssymb}
\usepackage{graphicx}
\usepackage {verbatim}

\usepackage{graphicx,amssymb,amstext,amsmath}
\usepackage{epsfig}
\usepackage{epstopdf}
\usepackage{color,graphics}

\DeclareMathOperator{\sech}{sech}

\begin{document}

\title{Coupled
circularly polarized electromagnetic soliton states in magnetized plasmas}
\author{G.P. Veldes}
\affiliation{Department of Physics, University of Thessaly, Lamia 35100, Greece}
\author{N. Lazarides}

\affiliation{Department of Mathematics, Khalifa University, 
Abu Dhabi, United Arab Emirates}

\author{D.J. Frantzeskakis}

\affiliation{Department of Physics, University of Athens, Panepistimiopolis,
Zografos, Athens 15784, Greece}
\author{I. Kourakis}

\affiliation{Department of Mathematics, Khalifa University, 
             Abu Dhabi, United Arab Emirates}

\affiliation{Space $\&$ Planetary Science Center, Khalifa University,
             Abu Dhabi, United Arab Emirates}

\affiliation{Visiting Researcher, Department of Physics, University of Athens,
             Panepistimiopolis, Zografos, Athens 15784, Greece}

\affiliation{Adjunct Researcher, Hellenic Space Center, Leoforos Kifissias 178,
             Chalandri, GR-15231 Athens, Greece}

\begin{abstract}
The interaction between two co-propagating electromagnetic pulses in a magnetized plasma is considered, from first principles, relying on a fluid-Maxwell model. Two circularly polarized wavepackets by same group velocities are considered, characterized by opposite circular polarization, to be identified as left-hand- or right hand circularly polarized (i.e. LCP or RCP, respectively). A multiscale perturbative technique is adopted, leading to a pair of coupled nonlinear Schr¨odinger-type (NLS) equations for the modulated amplitudes of the respective vector potentials associated with the two pulses. Systematic analysis reveals the existence, in certain frequency bands, of three different types of vector soliton modes: an LCP-bright/RCP-bright coupled soliton pair state, an LCP-bright/RCP-dark soliton pair, and an LCP-dark/RCP-bright soliton pair. 
The value of the magnetic field plays a critical role since it determines the type of vector solitons that may occur in certain frequency bands and, on the other hand, it affects the width of those frequency bands that are characterized by a specific type of vector soliton (type). The magnetic field (strength) thus arises as an order parameter, affecting the existence conditions of each type of solution (in the form of an envelope soliton pair).  
An exhaustive parametric investigation is presented in terms of frequency bands and in a wide range of magnetic field (strength) values, leading to results that may be applicable in beam-plasma interaction scenarios as well as in space plasmas and in the ionosphere.
\end{abstract}


\maketitle

\section{Introduction }

Electromagnetic (EM) pulse propagation in a plasma has been in the focus of
researchers for decades. From a basic theoretical point of view, it is
associated with fundamental physical processes involving various nonlinear
mechanisms and instabilities, which have been attracting increasing interest
since the early days of plasma research.
Thanks to a series of seminal early studies \cite{Akhiezer,Kozlov,Kaw,Esirkepov}
which set the foundations of contemporary laser-plasma interaction science
\cite{Kruer,Eliezer}, the study of electromagnetic pulse propagation has been
shown to be relevant with phenomena like parametric instabilities \cite{Quesnel,Gleixner},
harmonic generation \cite{Mori1}, self-focusing \cite{Esarey1}, intense electric
and magnetic field \cite{Borghesi1,Wagner}, and wakefield \cite{Chen,Roy} 
generation, among others. 
As laser power and technological sophistication increased, in the last decades, 
new regimes were attained where ultrashort ultrastrong electromagnetic (laser) 
pulses could be realized in the laboratory, thus entering the realm of relativistic 
dynamics (where the electron quiver velocity is comparable to the speed of light) 
\cite{Melrose-book} and, more recently, even quantum-electrodynamics  
\cite{DiPiazza,Brodin,Nargesi,Bulanov2020}. 
EM beam propagation is nowadays a cutting edge topic in science, due to its 
relevance with applications including laser-assisted (inertial confinement fusion, ICF) 
schemes for energy production, to ion acceleration \cite{Macchi1} and to 
sophisticated experimental diagnostic techniques \cite{Borghesi-review-LPB} 
and thus one of the hottest topics in contemporary research.

The propagation of relativistic soliton pulses,
which is our focus here, has not only been predicted by analytical theory
\cite{Kozlov,Kaw,Esirkepov,Akhiezer,marburger1975,Feng} and computer simulations
\cite{Bulanov_92,bulanov1999,Sentoku,Bulanov1,naumova2001,Esirkepov_02,bulanov2006,wu2013},
but also observed in laser-plasma interaction (LPI) experiments
\cite{borghesi2002,romagnani2010,sarri2010,sarri2011,Chen_07,Pirozhkov_07,Sylla_12,Blackman}.
These are essentially electromagnetic pulses trapped in a plasma density
cavitation with over-dense boundaries \cite{Sen,Mima,Farina}.
As shown in Refs. \onlinecite{Sentoku} and \onlinecite{Bulanov1}, a big portion 
of the laser pulse (electromagnetic) energy becomes localized into a train 
of soliton-like structures, which may play a crucial role in laser-matter and 
laser-plasma interaction processes.

The generation of relativistic electromagnetic solitons is governed by various 
physical mechanisms, including dispersion due to the finite electron inertia 
and nonlinearity due to a relativistic electron mass variation as well as to 
ponderomotive forces. Various analytical studies have focused on elucidating 
the properties of relativistic EM solitons, most of them focusing on a 
one-dimensional (1D) relativistic fluid model comprising of coupled
nonlinear equations for the vector and scalar potentials, coupled to the plasma. 
Circularly polarized (CP) EM solitons in a cold plasma were investigated early 
on by Kozlov \emph{et al} \cite{Kozlov}, who relied on a quasineutrality 
hypothesis to establish the occurrence of small-amplitude localized solutions 
(solitary waves). Later, computer simulations have shed some light on 
large-amplitude (relativistic) pulses, wherein charge separation in the plasma 
may be significant \cite{Kozlov}. Drifting envelope solitons of CP light were 
later modeled by Kaw and coworkers \cite{Kaw}. 
A stationary (zero group velocity) relativistic EM soliton solution with was shown 
to exist within a 1D cold plasma model in Ref. ~\onlinecite{Esirkepov}. 
Further remarkable works on 1D solitary waves in cold plasmas 
based on a perturbation technique include the studies by Kuehl and Zhang \cite{Kuehl} 
and by Farina \textit{et al} \cite{Farina,Farina2}, who incorporated the effect of ion motion in a fluid-Maxwell description. Later,  Poornakala \textit{et al}
\cite{Poornakala} discussed the existence of bright or dark type envelope 
solitons and identified associated regions in parameter space. Thermal 
effects were taken into account in Refs. \onlinecite{Lontano} and \onlinecite{Eliasson}. 
Single peak and multi-peak structures and their stability and mutual interaction properties
were studied by Saxena \emph{et al}, considering a non-uniform plasma \cite{Saxena1}, 
and also later in Refs. \onlinecite{Saxena2} and \onlinecite{Lehmann}. Note, for rigor, 
that the aforementioned studies have focused on non-magnetized plasma.

Magnetization 
during laser plasma interactions has attracted significant attention, 
as it impacts both the laser pulse dynamics and the substrate i.e. the background  
plasma properties. Inertial confinement fusion (ICF) has triggered interest in the 
study of magnetized plasma-laser interaction 
\cite{Pukhov1,Hurricane}. 
Experimental observations have confirmed the generation of intense magnetic 
fields (up to hundreds of MG) \cite{Borghesi3,Briand,Wagner,Tatarakis} during LPI. 

As regards EM pulse propagation in magnetized plasma, Shukla and Stenflo 
\cite{Shukla1} studied stationary solitary wave electromagnetic field structures 
adopting a slowly varying envelope approximation, followed by 
Nagesha \emph{et al} \cite{Nagesha}, who adopted a slowly varying envelop 
approximation and neglected perturbations in the
longitudinal component of the electron momentum, to find stationary solutions
for CP EM waves in cold magnetized plasma. Weakly super-acoustic (supersonic) 
EM envelope waves were studied by Rao \cite{Rao}. 
The occurrence of standing 1D relativistic solitons in cold magnetized plasmas
and the magnetic field's role on soliton stability was discussed by Farina
\emph{et al} \cite{Farina3}, who found that the frequency interval for stable 
solitons depends on both the magnitude and  the orientation of the
magnetic field in a significant manner. Furthermore, the maximum field 
amplitude of the solitonic pulses essentially depends on the ambient magnetic field (strength).

In the article at hand, we have undertaken an analytical and numerical study of 
the interaction between left- and right-hand circularly polarized  solitons in 
magnetized plasmas. Based on a fluid-Maxwell model, as starting point, 
we have established a system of scalar equations governing the propagation of 
circularly polarized EM waves in magnetized plasma.
First, analysis of the linear regime has shown that, in certain frequency bands,
left-hand circularly polarized (LCP-) and right-hand circularly polarized (RCP-)
modes can propagate at the same group velocity. Next, we adopt a large-amplitude
(nonlinear) approach, treating the model equations within an asymptotic
multiscale expansion method. Assuming the existence of coupled localized
(soliton) states for the magnetic vector potential, the analysis leads to a
system of two coupled nonlinear Schr{\"o}dinger (NLS) type equations for the
respective EM field envelopes, i.e. for the right-hand circularly polarized (RCP-)
and the left-hand circularly polarized (LCP-) mode respectively.
The analysis of the coupled NLS equations shows that there are three types coupled 
(vector) solitons for certain frequency bands, namely,\\
(a) a RCP-propagating bright soliton coupled with a LCP-propagating bright
    soliton,\\
(b) a RCP-propagating bright soliton coupled with a LCP-propagating dark soliton,
    and \\
(c) a RCP-propagating dark soliton coupled with a LCP-propagating bright soliton.

This paper is organized as follows. In Section II, we introduce the fluid-Maxwell
model for magnetized plasma and discuss the underlying physics. In Section III,
the system is linearized and the dispersive modes obtained are identified.
In Section IV, we assume a coupled-mode solution, and apply a multiscale
perturbative technique to derive a system of (two) coupled NLS equations for the
respective wavepacket amplitudes. In Section V, we present analytical results for
each type of vector soliton. Finally, in the concluding  Section VI, we summarize
our results.

\section{Fluid-Maxwell model for EM waves}

We consider an electron-ion plasma, permeated by a uniform magnetic field along
$\hat x$. The instantaneous state of the plasma is given by a number of dynamical
state variables, namely: the electron number density $n$, fluid speed
$\mathbf{u} = u \, \hat x$ and momentum (vector) $\mathbf{p} = (p_y, p_z)$.
These are subject to the influence of the electrostatic (scalar) potential
$\phi$ and the magnetic (vector) potential $\mathbf{A} = (A_y, A_z)$. All of
these variables are functions of space $x$ and time $t$, where one-dimensional
(1D) propagation was considered, in the direction parallel to the magnetic field,
i.e. along $\hat x$. Given the (high) frequency of interest, the ions will be
assumed to be stationary.

Our analysis will be based on the following closed system of scalar equations
which govern the propagation of the circularly polarized EM waves in magnetized
plasma:
\begin{eqnarray}
&&\frac{\partial^2 A_y}{\partial x^2}-\frac{\partial^2 A_y}{\partial t^2}=\frac{n}{\gamma}p_y,
\label{eq:eq1} \\
&&\frac{\partial^2 A_z}{\partial x^2}-\frac{\partial^2 A_z}{\partial t^2}=\frac{n}{\gamma}p_z,
\label{eq:eq2} \\
&&\frac{\partial }{\partial t}(p_y-A_y)+u\frac{\partial }{\partial x}(p_y-A_y)=-\frac{\Omega p_z}{\gamma}
\label{eq:eq3} \\
&&\frac{\partial }{\partial t}(p_z-A_z)+u\frac{\partial }{\partial x}(p_z-A_z)=\frac{\Omega p_y}{\gamma}
\label{eq:eq4} \\
&&\frac{\partial n}{\partial t}+\frac{\partial (nu)}{\partial x}=0,
\label{eq:eq5} \\
&&\frac{\partial (\gamma u)}{\partial t}=\frac{\partial} {\partial x}(\phi-\gamma)\nonumber\\
&&+ \frac{1}{\gamma}\left[ p_y\frac{\partial} {\partial x}(p_y-A_y)
+p_z\frac{\partial} {\partial x}(p_z-A_z)\right]\\
\label{eq:eq6}
&&\frac{\partial^2 \phi}{\partial x^2}=n-1 .
\label{eq:eq7}
\end{eqnarray}
All physical quantities have been normalized by appropriate scales, namely:
the scalar and vector potentials are normalized by $mc^{2}/e$, the electric
field $\mathbf{E}$ by $m c \omega_{pe}/e$, the magnetic field by $\mathbf{B}$ by
$m \omega_{pe}/e$, the momentum by $mc$, the density by the $n_{0}$, the electron
velocity by the light velocity $c$. Thus, in this framework the length is
measured in units of the skin length $c/\omega_{p0}$, and time is measured in
units of the plasma period (inverse plasma frequency) $(\omega_{pe}^{-1})$, where
$\omega_{pe}=\sqrt{4\pi n_{0} e^{2}/ m_{e}}$ (here $n_{0}$ denotes the
equilibrium electron density). Also, the electron momentum can be expressed as ,
\begin{equation}
\mathbf{P}=\mathbf{p}(x,t) + \gamma u(x,t) \hat x \label{P}
\end{equation}

for a circularly polarized EM pulse, where $\gamma$ is the relativistic factor 
and we have used the notation $\alpha=+1$
($\alpha=-1$) for left- (right-) hand circularly polarized electromagnetic waves.

\section{Multiple scale analysis for a single circularly polarized monochromatic
wavepacket}

Let the state vector $\textbf{S}=(n, u, \phi; A_y, A_z; p_y, p_z)$ describe the
state of the system at a given position $x$ and time $t$. Assuming a moderate
(but finite) deviation from the equilibrium state $S^{(0)}=(1,0,0; 0,0; 0,0)$,
one may adopt a multiple scales perturbation technique by expanding the system's
state as a series in terms of powers of a small parameter $\epsilon\ll 1$ and
subsequently considering multiharmonic analysis in each order. The method was
described exhaustively in the past, e.g. in Ref. \onlinecite{Jeffrey} in general,
and also, for electrostatic and electromagnetic waves in plasmas, in Ref.
\onlinecite{IKNPG} and \onlinecite{IKGVrogue} respectively, hence lengthy details
will be omitted here.

The system's state $\textbf{S}$ is expanded as
\begin{eqnarray}
\textbf{S} &=& \textbf{S}^{(0)}+ \epsilon \textbf{S}^{(1)}+ \epsilon^2 \textbf{S}^{(2)}+ ... \nonumber \\
&=& \textbf{S}^{(0)}+\sum_{n = 1}^{\infty}\epsilon^{n} \textbf{S}^{(n)}
\label{eq:ansatz01}
\end{eqnarray}
where $S^{(0)}=(1,0,0; 0,0; 0,0)$ demonstrates the equilibrium state of the 
system and $\epsilon\ll 1$ is a dimensionless, real and small parameter.

We now introduce new independent spatial and temporal variables, 
$x_m = \epsilon^m x$, $t_m = \epsilon^m t$ ($m=0,1,2,\cdots$), and accordingly expand
the space and time derivative operators $\partial_x = \partial_{x_0} + \epsilon \partial_{x_1} + \ldots$ 
and $\partial_t = \partial_{t_0} + \epsilon \partial_{t_1} + \ldots$.
Now, we use the assumption that the wavenumber $k$ and the frequency $\omega$ 
(fast variables) affect the perturbed state only via the phase  $kx-\omega t$ of 
the pulse, which are normalized by $\omega_{pe}$ and $k_{pe}=\omega/c$, respectively.
On the other hand, the amplitude of various frequency harmonics is assumed to be 
a slowly varying function in space and time, and will thus assumed to vary only on 
the slower scales ($x_2, x_3, ...; t_2, t_3, ...$).
The  solutions to be sought of Eqs.~(\ref{eq:eq1}-\ref{eq:eq7}) are therefore expressed in the form:
\begin{equation}
\textbf{S}^{(n)} =\sum_{\ell =-\infty}^{\ell=\infty}  \textbf{S}^{(n\ell)}(x_{m \geq 1}, t_{m \geq 1}) \, e^{i \ell (kx-\omega t)}.
\label{eq:ansatz02}
\end{equation}
%
The methodology adopted in this work relies on a Newell type multiscale technique, well known from nonlinear optics (where, interestingly space and time are reversed, hence dispersion occurs in the frequency domain, rather than the wavenumber space). The fundamental aim of the method it to allow for a slow variation of the wavepacket’s amplitude (envelope) in space and time. An excitation is considered around the equilibrium state (represented by the zeroth order state in a smallness parameter ($\epsilon^0$), while the linear (harmonic) carrier wave regime is obtained in first order ($\sim \epsilon^1$) and the group velocity frame is defined in the second order ($\sim \epsilon^2$), upon imposing certain compatibility conditions (namely, dictating the annihilation of secular terms occurring in that order). In each order in $\epsilon$, a secondary expansion is considered in harmonics, from the zeroth to the n-th harmonic. Within this (tedious but straightforward) scheme, the first harmonic (carrier) amplitudes arise as solutions in the first order, while the second harmonic amplitudes will arise via the second-order second-harmonic amplitudes. Zeroth-order amplitudes will originate from the 2nd order 2nd harmonic expressions, in combination with their 3rd order counterparts. Finally, an evolution equation is obtained for the envelope, by eliminating the secular terms occurring in 3rd order (1st harmonics). Linear analysis is therefore embedded in this analysis, thus generalized to account for the generation of higher harmonics.

Substituting  Eqs. (\ref{eq:ansatz01})-(\ref{eq:ansatz02}) into
Eqs. (\ref{eq:eq1})-(-\ref{eq:eq7})) and collecting the terms arising in each
order in $\epsilon$, we obtain the amplitude evolution equations at successive
orders.

In order ${\cal O} (\epsilon$) (linear limit), we have the equations
\begin{eqnarray}
&&\partial_{t_0}n^{(1)}+\partial_{x_0} u^{(1)}=0,
\label{eq:pl1} \\
&&\partial_{t_0} u^{(1)}-\partial_{x_0} \phi^{(1)}=0,
\label{eq:pl2}\\
&&\partial_{x_0}^2 \phi^{(1)}-n^{(1)}=0,
\label{eq:pl3}\\
&&(\partial_{x_0}^2 -\partial_{t_0}^2)A_{y}^{(1)}-p_{y}^{(1)}=0,
\label{eq:f1}\\
&&(\partial_{x_0}^2 -\partial_{t_0}^2)A_{z}^{(1)}-p_{z}^{(1)}=0,
\label{eq:f2}\\
&&\partial_{x_0}(p_{y}^{(1)}-A_{y}^{(1)})+\Omega p_{z}^{(1)}=0,
\label{eq:f3}\\
&&\partial_{x_0}(p_{z}^{(1)}-A_{z}^{(1)})-\Omega p_{y}^{(1)}=0.
\label{eq:f4}
\end{eqnarray}
%
\begin{itemize}
\item for the zeroth harmonic ($\ell=0$)
\begin{eqnarray}
&&n^{(10)}=u^{(10)}=0,\\
&&p_y^{(10)}=p_z^{(10)}=0,\\
&&A_y^{(10)}=A_z^{(10)}=0.
\label{eq:zero_harm}
\end{eqnarray}
\item  for the first harmonic ($\ell=1$)
\begin{eqnarray}
&& n^{(11)}=u^{(11)}=0,
\label{eq:first_harm1}\\
&&(\omega^2-k^2) A_y^{(11)} = p_y^{(11)},
\label{eq:eq11} \\
&&(\omega^2-k^2) A_z^{(11)}= p_z^{(11)},
\label{eq:eq21} \\
&&-i\omega(p_y^{(11)}-A_y^{(11)})= -\Omega p_z^{(11)},
\label{eq:eq31} \\
&&-i\omega(p_z^{(11)}-A_z^{(11)})=\Omega p_y^{(11)}.
\label{eq:eq41}
\end{eqnarray}
\end{itemize}
%
Considering the first harmonic amplitudes ($\ell=1$),
Eqs. (\ref{eq:eq11})-(\ref{eq:eq41}) require an explicit compatibility condition
to be adopted, in the form
\begin{equation}
\omega^2-k^2=\frac{\omega}{\omega-\alpha\Omega}.
\label{eq:linear_disp}
\end{equation}
One recognizes the dispersion relation for circularly-polarized waves
\cite{Swanson}. The solution thus obtained for the first-harmonic amplitudes
reads:
\begin{eqnarray}
&&p_z^{(11)}=i\alpha p^{(11)}, \quad p^{(11)}\equiv p_y^{(11)}.
\label{eq:pz}\\
&&A_z^{(11)}=i\alpha A^{(11)}, \quad A^{(11)}\equiv A_y^{(11)}.
\label{eq:Az}
\end{eqnarray}
%
In order ${\cal O} (\epsilon^2$), we have the equations
\begin{eqnarray}
&&\partial_{t_0}n^{(2)}+\partial_{x_0} u^{(2)}\nonumber\\
&&=-\partial_{t_1}n^{(1)}-\partial_{x_0} (n^{(1)}u^{(1)})-\partial_{x_1} u^{(1)},\\
\label{eq:pl21} \nonumber\\
&&\partial_{t_0} u^{(2)}-\partial_{x_0} \phi^{(2)}\nonumber\\
&&=-\partial_{t_1}u^{(1)}+\partial_{x_1} (\phi^{(1)}+p_{y}^{(1)}\partial_{x_0}(p_{y}^{(1)}-A_{y}^{(1)})\nonumber\\
&&+p_{z}^{(1)}\partial_{x_0}(p_{z}^{(1)}-A_{z}^{(1)}),\\
\label{eq:pl22}\nonumber\\
&&\partial_{x_0}^2 \phi^{(2)}-n^{(2)}=-2\partial_{x_0}\partial_{x_1}\phi^{(1)},\\
\label{eq:pl23}\nonumber\\
&&(\partial_{x_0}^2 -\partial_{t_0}^2)A_{y}^{(2)}-p_{y}^{(2)}\nonumber\\
&&=-2(\partial_{x_0}\partial_{x_1}-\partial_{t_0}\partial_{t_1})A_{y}^{(1)}+n^{(1)}p_{y}^{(1)},\\
\label{eq:f21}\nonumber\\
&&(\partial_{x_0}^2 -\partial_{t_0}^2)A_{z}^{(2)}-p_{z}^{(2)}\nonumber\\
&&=-2(\partial_{x_0}\partial_{x_1}-\partial_{t_0}\partial_{t_1})A_{z}^{(1)}+n^{(1)}p_{z}^{(1)},\\
\label{eq:f22}\nonumber\\
&&\partial_{t_0}(p_{y}^{(2)}-A_{y}^{(2)})+\Omega p_{z}^{(2)}\nonumber\\
&&=-(\partial_{t_1}+u^{(1)}\partial_{x_0})(p_{y}^{(1)}-A_{y}^{(1)}),\\
\label{eq:f23}\nonumber\\
&&\partial_{t_0}(p_{z}^{(2)}-A_{z}^{(2)})-\Omega p_{z}^{(2)}\nonumber\\
&&=-(\partial_{t_1}+u^{(1)}\partial_{x_0})(p_{z}^{(1)}-A_{z}^{(1)}).\\
\label{eq:f24}\nonumber
\end{eqnarray}
Same as in the Ref. \cite {IKGVrogue},  in the first harmonic (($\ell=1$), the condition for annihilation of secular 
terms leads to $\partial \cdot /\partial t_1 + v_g \, \partial \cdot /\partial x_1 = 0$ 
(for $ \cdot = p_{y/z}^{(1)}$ or $A_{y/z}^{(1)}$), implying that the envelope 
moves at the group velocity $v_{g} = \omega'(k)$, given by:
\begin{equation}
v_{g}=
\frac{2k(\omega-\alpha \Omega)^2}{2\omega(\omega-\alpha \Omega)^2+\alpha \Omega} \, ,
\label{eq:gvelqc}
\end{equation}
Following the same procedure in order ${\cal O} (\epsilon^3$), 
and using the variables $X=x_1-v_g t_1 \equiv \epsilon(x-v_g t)$ and 
$T=t_2 \equiv \epsilon^2 t$, we derive from the non-secularity condition at
$O(\epsilon ^{3})$ the NLS Eq.~(\ref{eq:nls0})
\begin{equation}
 i \frac{\partial \psi}{\partial \tau} + P \frac{\partial^2 \psi}{\partial \xi^2}  + Q |\psi|^2 =0,
\label{eq:nls0} \\
\end{equation}
where $\psi$ denotes the amplitude $A_y^{(11)}$, the (slow) time and space 
variables are $\tau = t_2$ and $\xi = x_1 - v_g t_1$, and the dispersion coefficient 
$P$ and the nonlinear self-phase modulation (SPM) coefficient $Q$ are respectively 
given by the (real) expressions:
\begin{eqnarray}
 && P \equiv\frac{1}{2}\frac{\partial^{2}\omega}{\partial k^{2}}=
\frac{v_{g}}{2k}+\frac{v_{g}^2}{\omega-\alpha\Omega}+\frac{v_{g}^3(3\omega-\alpha\Omega)}{2k(\omega-\alpha\Omega)}
\label{eq:dispcoef10} \\
&& Q=\frac{v_g}{k} (\omega^{2}-k^{2})^4 \, .
\label{eq:nlcoef10}
\end{eqnarray}
\begin{figure}
\includegraphics[scale=0.5]{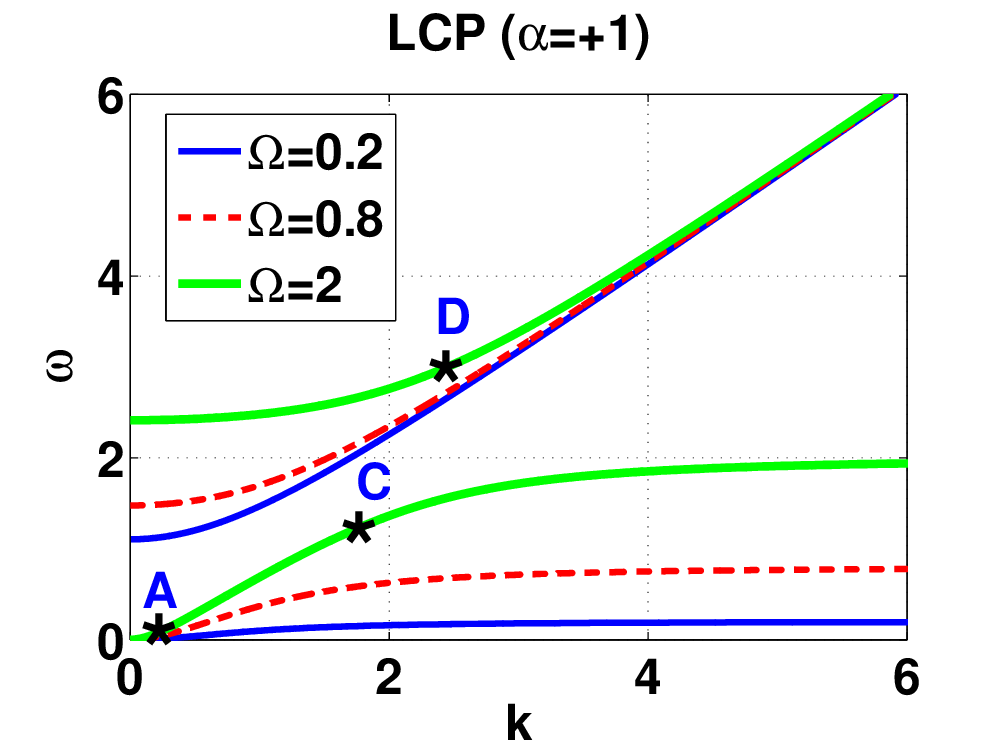}
\includegraphics[scale=0.5]{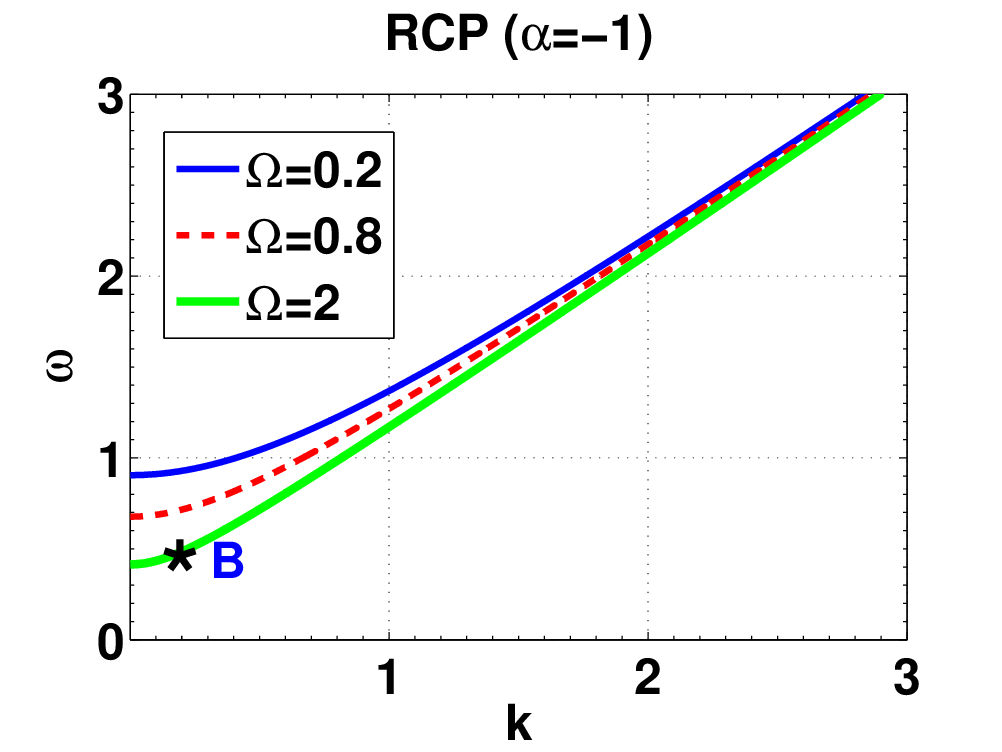}
\caption{The dispersion relation showing the  frequency $\omega$ as a
function of the  wave number $k$ (normalized values).  
The different values of $\Omega$, i.e., $\Omega=2$, $\Omega=0.8$, and 
$\Omega=0.2$ are depicted by bold (green), dashed (red) and thin solid (blue) 
lines, respectively. Top panel: The  LCP EM waves. Bottom panel: The RCP EM waves.}
\label{fig:dispersion}
\end{figure}

\section{Multiple scale analysis for coupled modes}
Now, we consider for a solution in the form
\begin{equation}
S^{(11)}=\sum_{j=1}^{2}S_j^{(11)}(x_1,x_2,\ldots, t_1,t_2,\ldots)\exp(i\theta_j) +{\rm c.c.},
\label{eq:ansatzA}
\end{equation}
where subscripts $j=1$ and $j=2$ correspond to the LCP and RCP waves, $S_j^{11}$
is an unknown complex function, $\theta_j = k_j x_0-\omega_j t_0 $, while
the wavenumbers $k_j$ and frequencies $\omega_j$ satisfy the dispersion relation
provided in Eq.~(\ref{eq:linear_disp}).
In Fig.~\ref{fig:dispersion}, we show the dispersion relation for the  $\Omega=0.2$,
$\Omega=0.8$, and $\Omega=2$. It is clear, from Fig.~\ref{fig:dispersion}, that for LCP
EM wave (top panel) there exist two frequency bands where the propagation is
possible: the high-frequency band and the low-frequency band. For $\Omega=0.2$
(depicted by solid (blue line) the propagation is possible for $\omega>1.105$
(high-frequency band) and for $\omega<0.2$ (low-frequency band) namely a gap
appears between $0.2<\omega<1.105$ where EM wave propagation is not possible.
For $\Omega=0.8$ (depicted by dashed (red) line) the propagation is possible for
$\omega>1.477$ (high-frequency band) and for $\omega<0.8$ (low-frequency band)
where the gap appears between $0.8<\omega<1.477$. Also, for $\Omega=2$ (depicted
by thick (green) line) the propagation is possible for $\omega>2.414$
(high-frequency band) and for $\omega<2$ (low-frequency band). In this case the
gap appears between $2<\omega<2.414$.
Moreover, for the RCP EM wave (bottom panel) propagation is possible for
frequencies $\omega>0.9047$ when $\Omega=0.2$ (depicted by solid (blue line).
In case where $\Omega=0.8$(depicted by dashed (red) line) and $\Omega=2$
(depicted by thick (green) line) the propagation is possible for frequencies
$\omega>0.6767$  and $\omega>0.414$,respectively.

Furthermore, the gap which appears for LCP EM waves increases (decreases) as the
$\Omega$ decreases (increases). It is important to notice that there is a
frequency region, which is independent from  polarization of EM waves, where the
propagation cannot take place  for values of $\Omega<0.707$.

Below we will demonstrate that in the nonlinear setting, coupling between LCP
and RCP EM waves with equal group velocities is possible, and we consider the 
interaction between LCP and RCP waves. Observing Fig.~\ref{fig:dispersion} 
and knowing that the  group velocity at a given wavenumber represents a tangent (line) to the dispersion curve, 
it is possible to identify domains, with equal group velocities.  Thus, using 
Eq.~(\ref{eq:linear_disp}) we obtain the group velocity
$v_{g} = \partial \omega/\partial k$:
\[ 
v_{g}=
\frac{2k(\omega-\alpha \Omega)^2}{2\omega(\omega-\alpha \Omega)^2+\alpha \Omega} \, ,
\]
i.e. Eq. (\ref{eq:gvelqc}) above. 

\begin{figure}[tbp]
\includegraphics[scale=0.5]{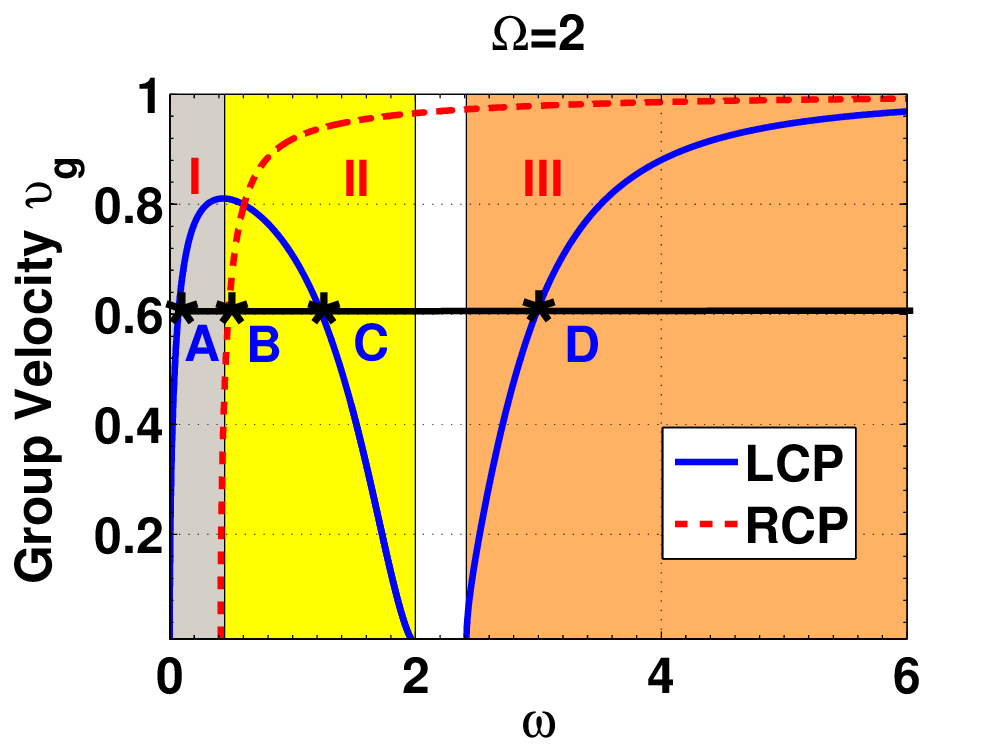}\\
\caption{The group velocity $\upsilon_g$ as a function of the  normalized frequency 
$\omega$. The dashed (red) and solid (blue) dashed (red) lines show the group velocity 
$\upsilon_g$ for RCP- and LCP- EM waves, respectively.}
\label{fig:HSgvel}
\end{figure}
This result is presented in Fig.~\ref{fig:HSgvel}, where the group velocity
$v_g$ is plotted as a function of the normalized frequency $\omega$.
Notice that the dashed (red) solid (blue) and solid (blue) lines show the group velocity
$\upsilon_g$  for RCP- and LCP- EM waves, respectively.
\begin{figure}[tbp]
\begin{tabular}{cc}
\includegraphics[scale=0.4]{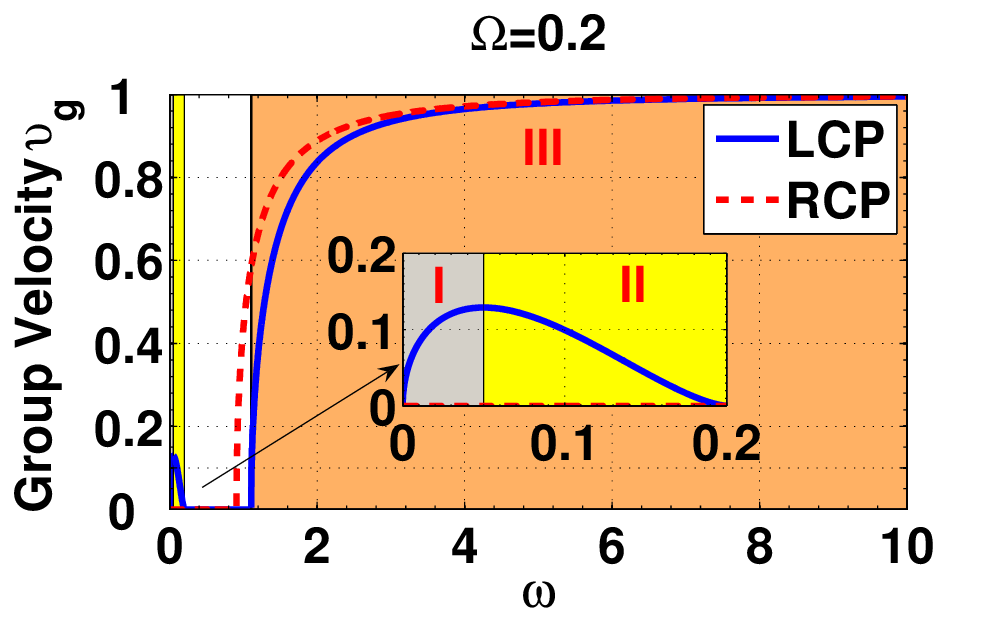}\\
\includegraphics[scale=0.4]{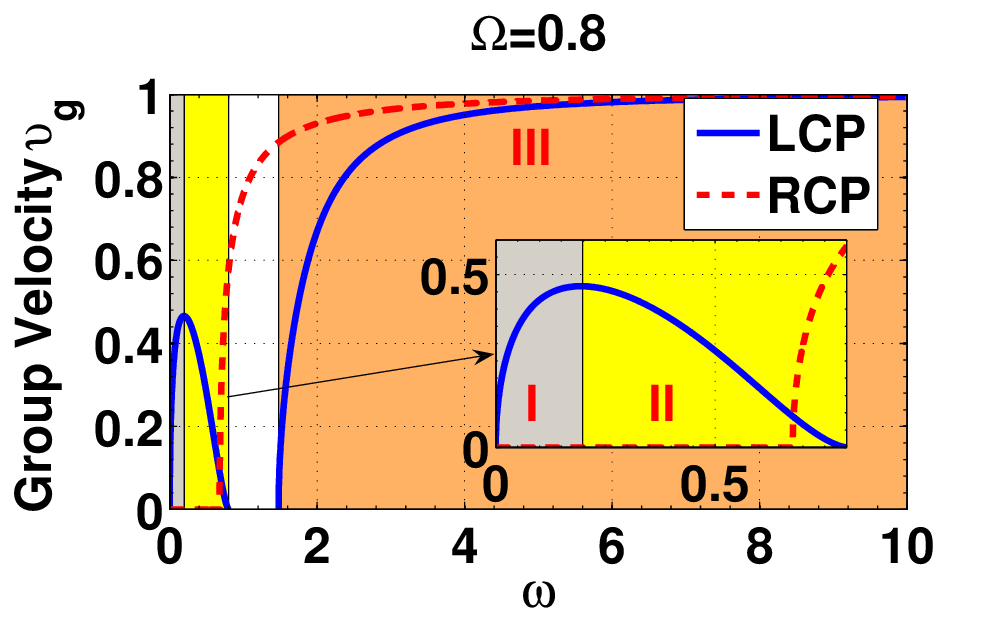}\\
\includegraphics[scale=0.4]{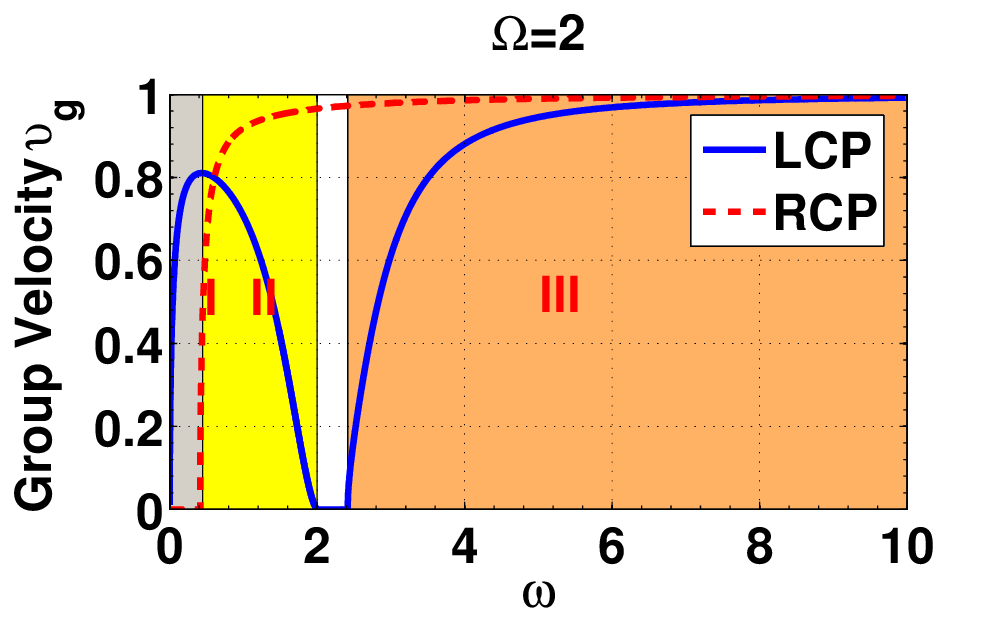}\\
\end{tabular}
\caption{The group velocity $\upsilon_g$ as a function of the  normalized
         frequency $\omega$ for different values of $\Omega$, i.e., $\Omega=0.2$
         (top panel), $\Omega=0.8$ (middle panel), and $\Omega=2$ (bottom panel).
         The dashed (red) and solid (blue)  dashed (red) lines show the group velocity
         $\upsilon_g$  for RCP- and LCP- EM waves, respectively.}
\label{fig:gvel}
\end{figure}

As seen in Fig.~\ref{fig:HSgvel} using a horizontal cut we can to observe 
that there are a RCP and a LCP electromagnetic wave which have the same group 
velocity (and are coupled  in the nonlinear regime).

The point A,B, C and D,
in Fig.~\ref{fig:HSgvel}, have the same group velocity, namely, $v_g=0.4$.
Also, we see that there is a maximum possible common  $v_{g_{max}}^{\ell}$  in
the low frequency band for LCP wave. Thus, one can divide each of the
group-velocity curves into three sub-regions, depending on the sign of the
group-velocity dispersion (GVD), $\partial v_g/\partial \omega$, where the 
interaction with equal group velocities may appear. As depicted in
Fig.~\ref{fig:HSgvel} these subregions are:
(a) the sub-bands I  and II  for the low frequency band of the LCP EM wave,
    characterized by positive and negative GVD respectively,
    and the sub-bands III for the high frequency band of the LCP EM wave,
    characterized by positive GVD,
(b) the band of the RCP EM wave, again characterized by positive GVD
    respectively.
Thus, nonlinear LCP and RCP modes of equal $v_g$ can feature the following three
different possible interactions:
\begin{itemize}
\item {\it Case 1}:band of the RCP-mode and  the LCP-mode in band I,
both featuring positive GVD for $v_g \leq v_{g_{max}}^{\ell}$.
\item {\it Case 2}: band of the RCP-mode   and  the LCP-mode in band II;
here, the RCP (LCP) mode features  positive (negative) GVD for  $v_g \leq v_{g_{max}}^{\ell}$.
\item {\it Case 3}: band of the RCP-mode   and  the LCP-mode in band III,
both featuring positive GVD for $0<v_g \leq 1$.
\end{itemize}

As shown in Fig.~\ref{fig:gvel}, upon inspection of the group-velocity curves 
for $\Omega=0.8$ (middle panel) we see that the maximum possible common  
$v_{g_{max}}^{\ell}=0.4661$  occurring at $\omega=0.198$, in the low frequency 
band for LCP wave. Then, the subregions are:
(a) the sub-bands I ($0<\omega<0.198$) and II ($0.198<\omega<0.8$) for the low 
    frequency band and the sub-bands III ($\omega>1.477$)for the high frequency 
    band of the LCP EM wave
(b) the band of the RCP EM wave ($\omega>0.6767$).

Notice that, $v_g=1$ is the group velocity for unmagnetized plasma.
It is obvious that the above considerations is the result of the existence 
of the frequency gap as shows in Fig. 3 for all values of $\Omega$, i.e, 
for $\Omega=0.2$, $\Omega=0.8  $and  $\Omega=2$.
%
Furthermore, we can observe that as $\Omega$ increases  the $v_g^{\ell}$ is raised 
and the gap is smaller. When the $\Omega\gg1$ then $v_g\rightarrow 1$ and the LCP 
and RCP wave are degenerated.

\subsection{Nonlinear analysis: the coupled NLS equations}
In our case, since we study the interaction between a LCP and a RCP nonlinear 
mode with equal group velocities, we seek for a solution of Eqs.~(\ref{eq:eq1}-\ref{eq:eq7}) in the form:
\begin{equation}
S^{(n)}=\sum_{j=1}^{2}S_j^{(n)}(X,T)\exp(i\theta_j) +{\rm c.c.},
\label{eq:ansatzB}
\end{equation}
where ``c.c.'' denotes complex conjugate. In Eq.~(\ref{eq:ansatzB}), subscripts 
$j=1,2$ correspond to the LCP and RCP mode, $S_j^{(n)}(X,T)$ are unknown 
(continuous) slowly-varying envelope functions depending on the slow scales $X=\epsilon(x-v_gt)$
(where $v_g$ is the {\it common} group velocity) and $T=\epsilon^2 t$, while $\exp(i\theta_{j})$,
with $\theta_{j} = k_j x-\omega_j t$, are the carriers of frequencies $\omega_j$ and
wavenumbers $k_j$.
The $\epsilon$ is a formal small parameter related to the soliton amplitude (see below).
%
According to Eq.~(\ref{eq:ansatzB}), the field $S_j^{(n)}$ is the leading-order form 
of a more general ansatz employing multiple time and space scales. In this context, 
use of a formal multi-scale expansion method leads to a hierarchy of equations at 
various powers of $\epsilon$, which are solved up to the third-order. Indeed, at orders 
$\mathcal{O}(\epsilon)$ (linear limit) and $\mathcal{O}(\epsilon^2)$, we derive the 
dispersion relation, Eq.~(\ref{eq:linear_disp}), and the group velocity, 
Eq.~(\ref{eq:gvelqc}), respectively. At the order, $\mathcal{O}(\epsilon^3)$
we obtain the  nonlinear coupled NLS equations as
%
\begin{eqnarray}
\!\!\!\!\!\!\!\!\!\!\!
&&i\partial_{T}\Psi_1+ D_1 \partial_{X}^{2}\Psi_1 + \left(g_{11}|\Psi_1|^2+ g_{12}|\Psi_2|^2\right)\Psi_1=0,
\label{eq:NLS1} \\
\!\!\!\!\!\!\!\!\!\!\!
&&i\partial_{T}\Psi_2+ D_2 \partial_{X}^{2}\Psi_2 + \left(g_{21}|\Psi_1|^2+ g_{22}|\Psi_2|^2\right)\Psi_2=0,
\label{eq:NLS2}
\end{eqnarray}
where the amplitude $\Psi_j$, the normalized GVD coefficients $D_j$, 
the self-phase modulation (SPM) coefficients $g_{jj}$, and
the cross-phase modulation (CPM) coefficients $g_{j,3-j}$ (with $j=1,2$) 
are respectively given by:
\begin{eqnarray}
&&\Psi_{j}\equiv A_j^{(11)}\\
 &&D_j\equiv\frac{1}{2}\frac{\partial^{2}\omega_j}{\partial k_j^{2}}=
\frac{v_{g}}{2k_j}+\frac{v_{g}^2}{\omega-\alpha\Omega}-\frac{v_{g}^3(3\omega-\alpha\Omega)}{2k_j(\omega-\alpha\Omega)}
\label{eq:dispcoef} \\
&&g_{jj}=\frac{v_g}{k_j} (\omega_j^{2}-k_j^{2})^4,
\label{eq:nlcoef1} \\
&&g_{j,3-j}=\frac{2v_g}{k_j} (\omega_j^{2}-k_j^{2})^2(\omega_{3-j}^{2}-k_{3-j}^{2})^2.
\label{eq:nlcoef2}
\end{eqnarray}
Note that, using the Eqs.~(\ref{eq:linear_disp}) and (\ref{eq:gvelqc}) the above relations be cast in the form:
\begin{eqnarray}
&&g_{jj}=\frac{2\omega_j^4}{[2\omega_j(\omega_j-\alpha \Omega)^2 + \alpha \Omega](\omega_j-\alpha \Omega)^2},
\label{eq:nlcoef11} \\
&&g_{j,3-j}=\frac{4\omega_j^2 \omega_{j,3-j}^2}{[2\omega_j(\omega_j-\alpha \Omega)^2 + \alpha \Omega](\omega_{j,3-j}-\alpha \Omega)^2}.
\label{eq:nlcoef22}
\end{eqnarray}
As shown from Eqs.~(\ref{eq:nlcoef1})-(\ref{eq:nlcoef2}) the coefficients $g_{jj}$, and
$g_{j,3-j}$ are always positive.
Next, using scale transformations, we measure normalized time $T$ and densities $|\Psi_{j}|^2$ in units
of $|D_1|^{-1}$ and $|D_1/g_{jj}|$ respectively, and cast Eqs.~(\ref{eq:NLS1})-(\ref{eq:NLS2}) in the form:
\begin{eqnarray}
&&i\partial_{T} \Psi_1+ s~\partial_{X}^2 \Psi_1 + \left(\Psi_1|^2+ \lambda_1|\Psi_2|^2\right)\Psi_1=0,
\label{eq:MNLS1} \\
&&i\partial_{T} \Psi_2+ d~\partial_{X}^2 \Psi_2 + \left(\lambda_2|\Psi_1|^2+ \Psi_2|^2\right)\Psi_2=0,
\label{eq:MNLS2}
\end{eqnarray}
where
\begin{eqnarray}
s&=&{\rm sign }(D_1), \quad d=\frac{D_2}{\left|D_1\right|},
\nonumber\\
\lambda_{1}&=&\frac{g_{12}}{|g_{22}|},\quad
\lambda_{2}=\frac{g_{21}}{|g_{11}|}.
\label{eq:intcoeff}
\end{eqnarray}
where the coefficients $\lambda_{1,2}$ and $d$ are positive.
As arises from Eqs.~(\ref{eq:MNLS1}) when the coupling coefficients ($\lambda_j=0$) 
are zero then the evolution of either the LCP wave $\Psi_1$ or the RCP wave $\Psi_2$ 
is described by a single NLS equation; the latter, supports soliton solutions of the 
dark or the bright type, depending on the relative signs of dispersion and 
nonlinearity coefficients (see, e.g., Ref.~\cite{yuri}). 
Particularly, the mode $\Psi_1$ supports dark solitons for $s<0$ or bright solitons 
for $s>0$. The mode $\Psi_2$ supports  only bright solitons because the coefficient 
$d$ is always positive.
However, because the above conditions are modified when $\lambda_j \ne 0$ various 
types of vector (coupled) solitons can be found as seen from the Eqs.~(\ref{eq:MNLS1}) 
and (\ref{eq:MNLS2}).

According to the Ref.~\citep{veldes}, four types of vector solitons are possible:
\begin{itemize}
\item {\it bright-bright} (BB) solitons, in the form:
\begin{eqnarray}
\Psi_1(X,T)&=&\Psi_{1,0}{\rm sech}(b X)\exp(-i\eta_1 T),
\label{eq:bsa}\\
\Psi_2(X,T)&=&\Psi_{2,0} {\rm sech}(b X)\exp(-i\eta_2 T).
\label{eq:bsb}
\end{eqnarray}
\item {\it bright-dark} (BD) solitons, in the form:
\begin{eqnarray}
\Psi_1(X,T)&=&\Psi_{1,0}{\rm sech}(b X)\exp(-i\eta_1 T),
\label{eq:bs2}\\
\Psi_2(X,T)&=&\Psi_{2,0} {\rm tanh}(b X)\exp(-i\eta_2 T),
\label{eq:ds2}
\end{eqnarray}
\item {\it dark-bright} (DB) solitons, in the form:
\begin{eqnarray}
\Psi_1(X,T)&=&\Psi_{1,0}{\rm tanh}(b X)\exp(-i\eta_1 T),
\label{eq:ds1}\\
\Psi_2(X,T)&=&\Psi_{2,0} {\rm sech}(b X)\exp(-i\eta_2 T),
\label{eq:bs1}
\end{eqnarray}
\item {\it dark-dark} (DD) solitons, in the form:
\begin{eqnarray}
\Psi_1(X,T)&=&\Psi_{1,0}{\rm tanh}(b X)\exp(-i\eta_1 T),
\label{eq:dd1s}\\
\Psi_2(X,T)&=&\Psi_{2,0} {\rm tanh}(b X)\exp(-i\eta_2 T).
\label{eq:dd2s}
\end{eqnarray}
\end{itemize}
In the above equations, $\eta_j$ (j=1,2) and  $\Psi_{j,0}$ declare the frequencies 
and amplitudes of each soliton, while $b$ is the (common) inverse width of the solitons.

Now, each of the above ansatz is substituted in Eqs.~(\ref{eq:MNLS1}) and (\ref{eq:MNLS2}), 
leading to a set of equations connecting the soliton parameters. Particularly, 
the equations connecting the amplitudes $\Psi_{j,0}$ and the inverse width $b$ are of the form:
\begin{eqnarray}
(\Psi_{2,0}/\Psi_{1,0})^2&=&-\alpha_1,\qquad (b/\Psi_{1,0})^2=-\alpha_2, \label{BB} 
\label{eq:sign_alpha1}\\
(\Psi_{2,0}/\Psi_{1,0})^2&=&\alpha_1,~~\qquad (b/\Psi_{1,0})^2=-\alpha_2, \label{BD}
 \label{eq:sign_alpha2}\\
(\Psi_{2,0}/\Psi_{1,0})^2&=&\alpha_1, ~~\qquad (b/\Psi_{1,0})^2=\alpha_2, \label{DB}
 \label{eq:sign_alpha3}\\
(\Psi_{2,0}/\Psi_{1,0})^2&=&-\alpha_1, \qquad (b/\Psi_{1,0})^2=\alpha_2, \label{DD}
\label{eq:sign_alpha4}
\end{eqnarray}
for the BB, BD, DB and DD solitons respectively, where parameters $\alpha_j$ ($j=1,2$) are given by:
\begin{equation}
\alpha_1=\frac{s \lambda_2 - d}{s-d\lambda_1},~ \alpha_2=\frac{\lambda_1 \lambda_2-1}{2(s-d \lambda_1)}.
\label{eq:coeff_alpha}
\end{equation}
\begin{figure*}[tbp]
\begin{tabular}{ccc}
&\includegraphics[scale=0.45]{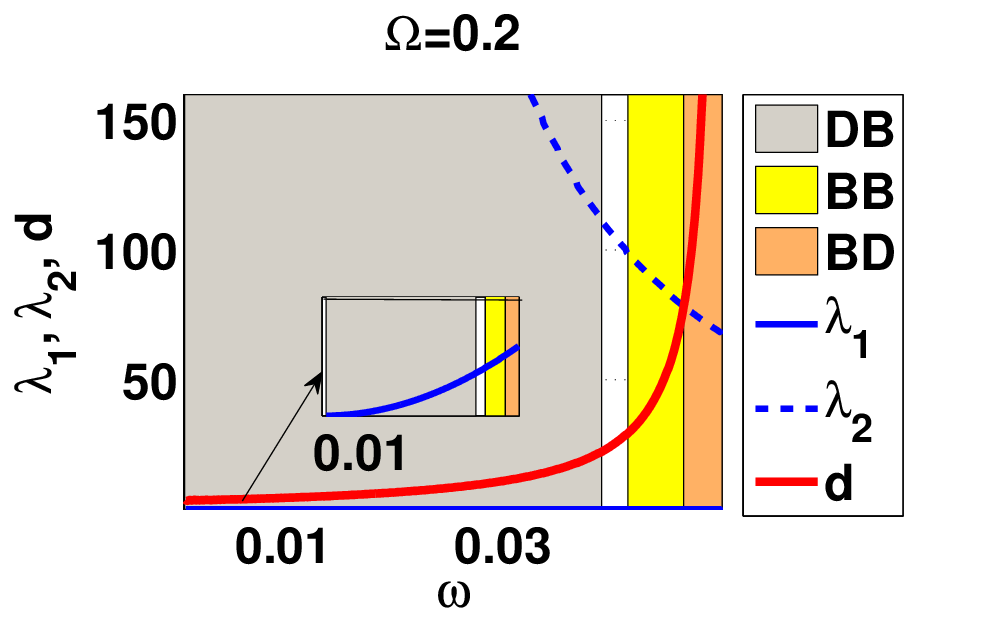}&\includegraphics[scale=0.45]{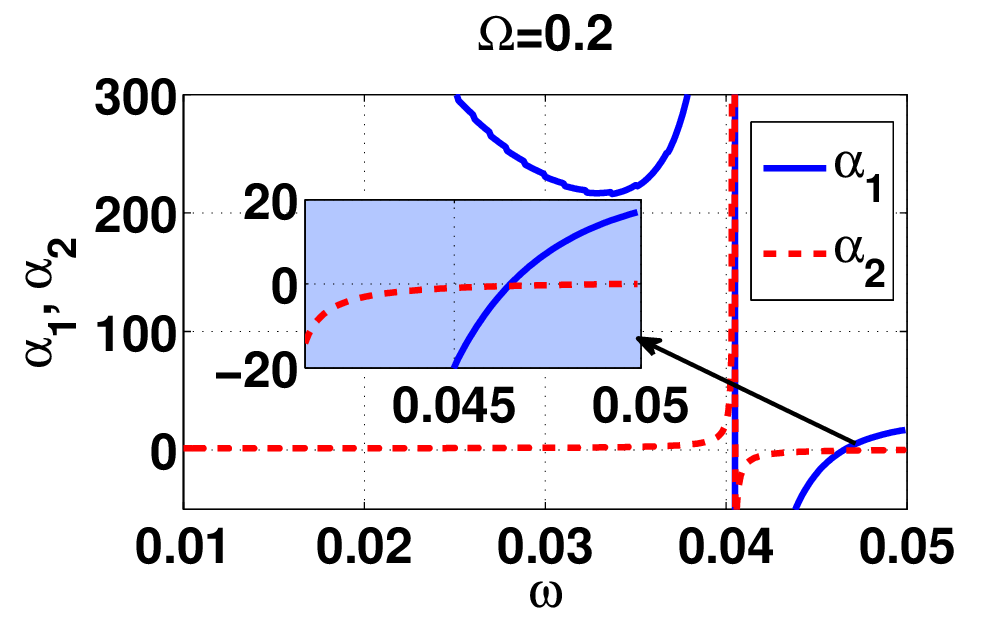}\\
&\includegraphics[scale=0.45]{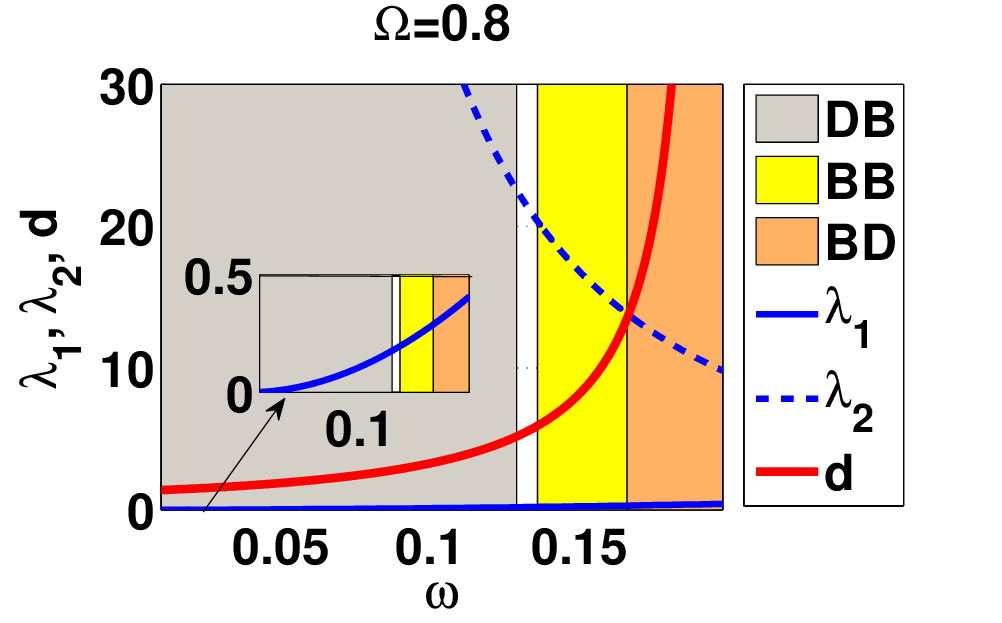}&\includegraphics[scale=0.45]{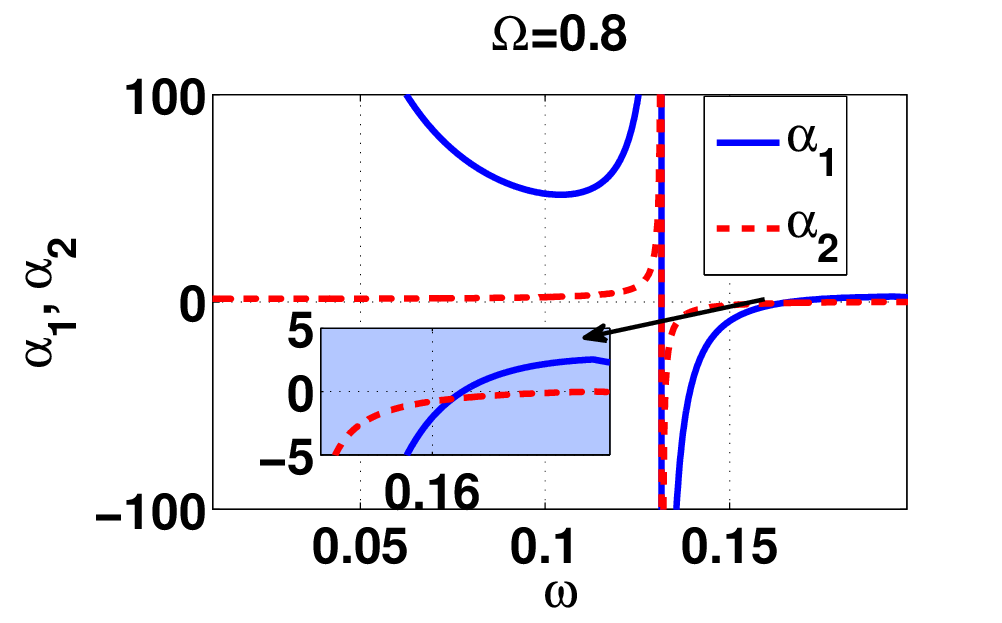}\\
&\includegraphics[scale=0.45]{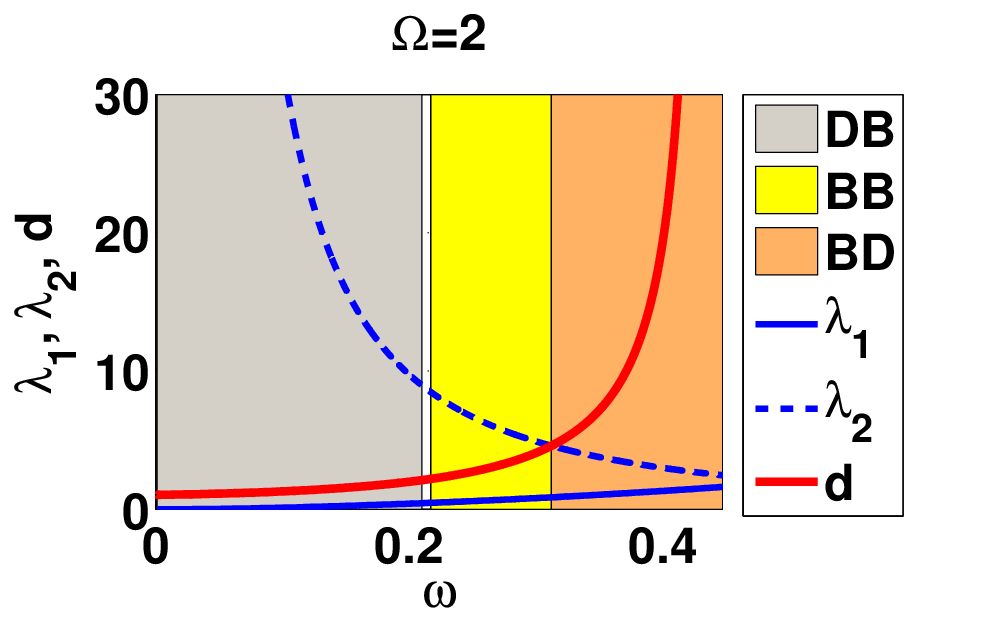}&\includegraphics[scale=0.45]{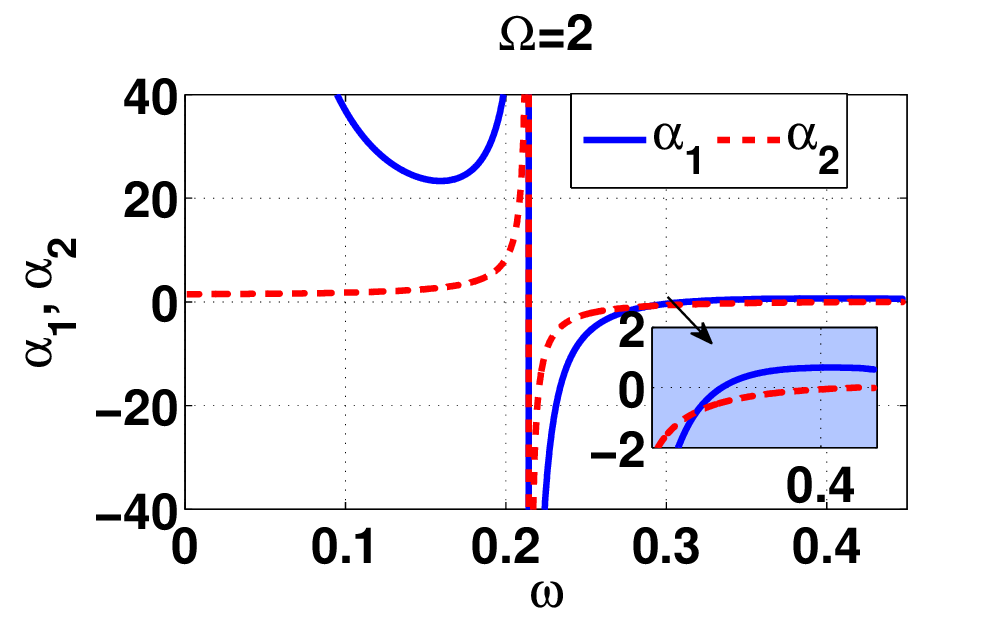}\\
\end{tabular}
\caption{Case 1: (Left panels) The dependence of the parameters, $\lambda_1$ [thin solid (blue) line], $\lambda_2$ [dashed (blue) line] and $d$ [bold solid (red) line] on the normalized frequency $\omega$,
for different values of $\Omega$, i.e., $\Omega=0.2$ (top panel), $\Omega=0.8$ (middle panel), and $\Omega=2$ (bottom panel).
(Right panels) The same for the coefficients  $\alpha_1$ [solid (blue) line], $\alpha_2$ [dashed (red) line].}
\label{fig:INTADcoeff}
\end{figure*}
%
It is worth noting that a similar set of coupled NLS equation has been derived
in the past for modulational wavepacket interaction in electron-ion plasmas,
considering either upper-hybrid \cite{Kourakis2005} or, more recently,
electrostatic \cite{NL1,NL2,NL3} waves. Although the mathematical setting is
analogous, and may therefore interest the reader of this paper, the physical
background in this paper is distinct from those works; in those earlier studies, for instance,
an unmagnetized plasma was considered, in the electrostatic approximation
(i.e. neglecting magnetization, thus precluding electromagnetic excitations),
with the focus being on the nonthermal (kappa-distributed) background electron
statistics (a common occurrence in Space plasmas). Contrary to that picture, 
our focus here is on electromagnetic solitary waves propagating in magnetized 
plasma and, in particular, on the role of the ambient magnetic field on their 
structural and propagation characteristics.

\section{Soliton interactions in different frequency bands}
%
It is important to understand that the sign of the parameters $\alpha_j$ ($j=1, 2$) 
defined above determines the type of vector solitons that may occur in the plasma, 
as dictated by  Eqs. (\ref{eq:sign_alpha1})-(\ref{eq:sign_alpha4}). Also, the parameters $\alpha_j$ depend on the frequency and the magnetic field according to Eq. (\ref{eq:coeff_alpha}), through coefficients $\lambda_{1,2}$, $s$ and $d$. Consequently, we expect that the precise solitons solutions will arise from the investigation of the sign of the parameters $\alpha_j$ and the value of magnetic field, for all cases (Case 1, 2 and 3) as have been defined in Section V.  On the other hand, the Eqs.~(\ref{eq:MNLS1})-(\ref{eq:MNLS2}) are no longer of the Manakov type and, thus, generally, they are not completely integrable. 
Nevertheless, standing wave soliton solutions whose exact analytical form depend  on by sign of  the parameters $\alpha_j$ can still be found. That is valid for all  cases of interactions (Case 1, 2,  and 3).

\subsection{Case 1:~Solitons in bands I and RCP.}

Initially, we study the coupling between a propagating soliton, 
with a frequency that is  in band  I, and a   propagating soliton, with a 
frequency that is in RCP band. In Fig.~\ref{fig:INTADcoeff} we depict the 
dependence of the parameters $\lambda_1$, $\lambda_2$ and $d$, as a function 
of the normalized frequency $\omega$ (for  $\Omega=0.2$, $\Omega=0.8$ and 
$\Omega=2$). Now, we have $s=+1$ (cf. Fig.~\ref{fig:gvel}) and $\lambda_{1,2}>0$ 
and $d>0$, as can be seen in the Fig.~\ref{fig:INTADcoeff}. 
Also, the parameters $\alpha_j$ ($j=1,2$) are given by:
\begin{eqnarray}
&&\alpha_1=\frac{\lambda_2-d}{1-d\lambda_1},
\label{eq:a1}\\
&&\alpha_2=\frac{\lambda_1 \lambda_2-1}{2(1-d\lambda_1)}.
\label{eq:a2}
\end{eqnarray}
In this case, the Eqs.~(\ref{eq:MNLS1})-(\ref{eq:MNLS2}) are no longer of the 
Manakov type and, thus, generally, they are not completely integrable. 
Nevertheless, standing wave solitons solutions can still be found  whose exact 
analytical form depend on by sign of  the parameters $\alpha_j$.
\textbf{1}.~ When $\alpha_1>0$ and $\alpha_2>0$ then we have a case of coupled 
solitons, a dark soliton and a bright soliton, whose exact analytical form are 
given by Eqs.~(\ref{eq:ds1})-(\ref{eq:bs1}), where the soliton amplitude parameters 
$\eta_{1,2}$ and the inverse width $b$ are connected via the following equations:
\begin{eqnarray}
&&\eta_1=-\Psi_{1,0}^2,
\label{eq:eta11}\\
&&\eta_2=-(d\alpha_2+\lambda_2)\Psi_{1,0}^2,
\label{eq:eta21}\\
&&b^2=\alpha_2 \Psi_{1,0}^2,
\label{eq:bb1}
\end{eqnarray}
It is thus clear that the above solutions have one free parameter. Employing the 
solutions (\ref{eq:ds1})-(\ref{eq:bs1}), we can approximate the vector potential 
$A_y(x,t)$, in terms of the original coordinates $x$ and $t$, as follows:
\begin{eqnarray}
A_{y}(x,t)& \approx & A_{y0}[\Psi_1(x,t) \cos(k_{1} x-\omega_{01}t)\nonumber\\
&+&\Psi_2(x,t) \cos(k_{2} x-\omega_{02}t)],
\label{eq:darkbright1}
\end{eqnarray}
The amplitude $A_0$ and the frequencies $\omega_{0j}$ ($j=1,2$) of the soliton are given by:
\begin{eqnarray}
A_{y0}&=&2\epsilon\sqrt{\left|\frac{D_1}{g_{11}}\right|},
\label{Ay01} \\
\omega_{0j}&=&\omega_j+ \epsilon^2\eta_j|D_1|.
\label{f01}
\end{eqnarray}
In this case,
\begin{eqnarray}
\!\!\!\!\!\!\!\!
&&\Psi_1=\Psi_{1,0}\tanh[\epsilon b(x-v_g t)],
\label{eq:dbsol11}\\
\!\!\!\!\!\!\!\!
&&\Psi_2=\sqrt{\left|\frac{g_{11}}{g_{22}}\right|\alpha_1}\Psi_{1,0}\sech[\epsilon b(x-v_g t)].
\label{eq:dbsol21}
\end{eqnarray}
\begin{figure}[tbp]
\begin{tabular}{cc}
\includegraphics[scale=0.4]{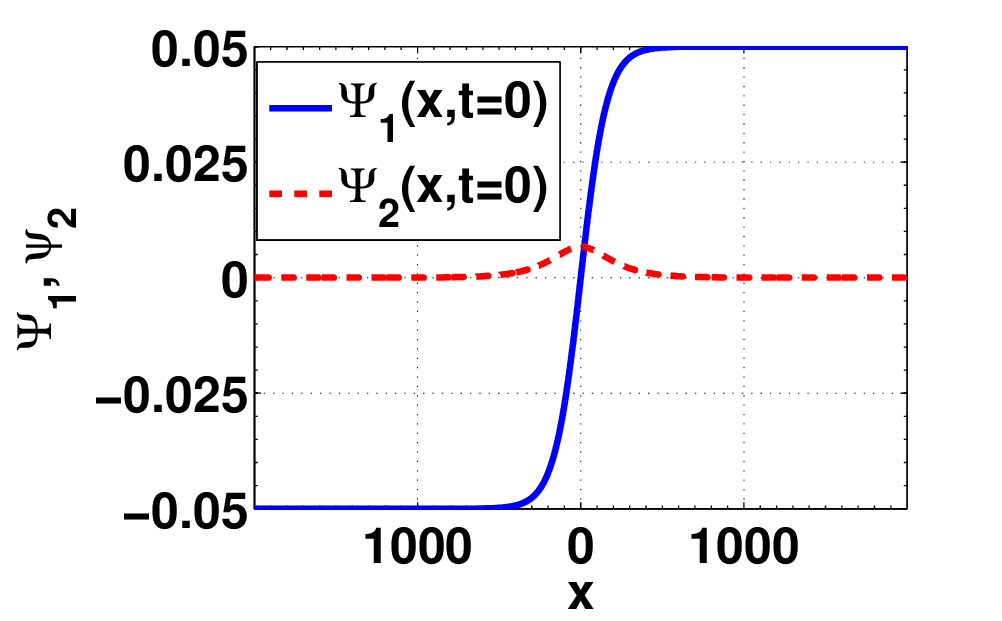}\\
\includegraphics[scale=0.4]{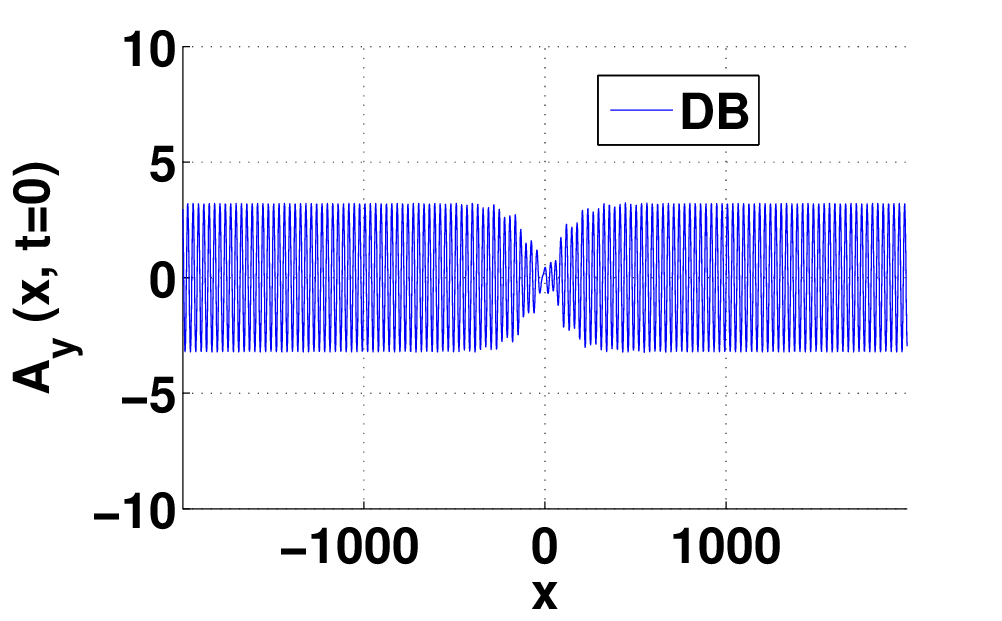}
\end{tabular}
\caption{Case 1: The profiles (at t=0) of the dark and bright solitons (top panel) 
and dark-bright soliton (bottom panel) for $\Omega=0.8$ . The parameters used are $v_g= 0.2965$,
$\omega_1=0.0335$ ($k_1=0.2119$) and $\omega_2= 0.7007$ ($k_2=0.1551$). Also, we use $\epsilon=0.1$ and $\Psi_{1,0}=0.05$.}
\label{fig:DB_sol}
\end{figure}

In Fig.~\ref{fig:DB_sol} we show the profiles (at t=0) of the dark (LCP mode)
and bright (RCP mode) solitons  in the absence of coupling (top panel) as well
as the dark-bright soliton (bottom panel), for $\Omega=0.8$.
Now, the group velocity of the DB soliton (common for both components) is
$v_g= 0.2965$, which occurs when the angular frequencies for the modes $\Psi_1$
and $\Psi_2$ take, respectively, the values $\omega_1=0.0335$ ($k_1=0.2119$) and
$\omega_2= 0.7007$ ($k_2=0.1551$).
\textbf{2}.~ When $\alpha_1<0$ and $\alpha_2<0$ then we have a case of coupled
bright solitons, whose exact analytical form are given by
Eqs.~(\ref{eq:bsa})-(\ref{eq:bsb}).
Now, the parameters $\eta_{1,2}$ and $b$ are connected via the following
equations:
\begin{eqnarray}
&&\eta_1=\alpha_2 \Psi_{1,0}^2,
\label{eq:eta1a}\\
&&\eta_2=d\alpha_2\Psi_{1,0}^2,
\label{eq:eta2b}\\
&&b^2=-\alpha_2 \Psi_{1,0}^2.
\label{eq:bbab}
\end{eqnarray}
The vector potential $A_y(x,t)$, in terms of the original coordinates $x$ and $t$ 
and soliton parameters, are given by the Eq.~(\ref{eq:darkbright1}) and 
Eqs.~(\ref{Ay01})-(\ref{f01}) respectively, where, now
\begin{eqnarray}
\!\!\!\!\!\!\!\!
&&\Psi_1=\Psi_{1,0}\sech[\epsilon b(x-v_g t)],
\label{eq:dbsol1a}\\
\!\!\!\!\!\!\!\!
&&\Psi_2=\sqrt{\left|\frac{g_{11}}{g_{22}}\alpha_1\right|}\Psi_{1,0}\sech[\epsilon b(x-v_g t)].
\label{eq:dbsol2b}
\end{eqnarray}
\begin{figure}[tbp]
\begin{tabular}{cc}
\includegraphics[scale=0.4]{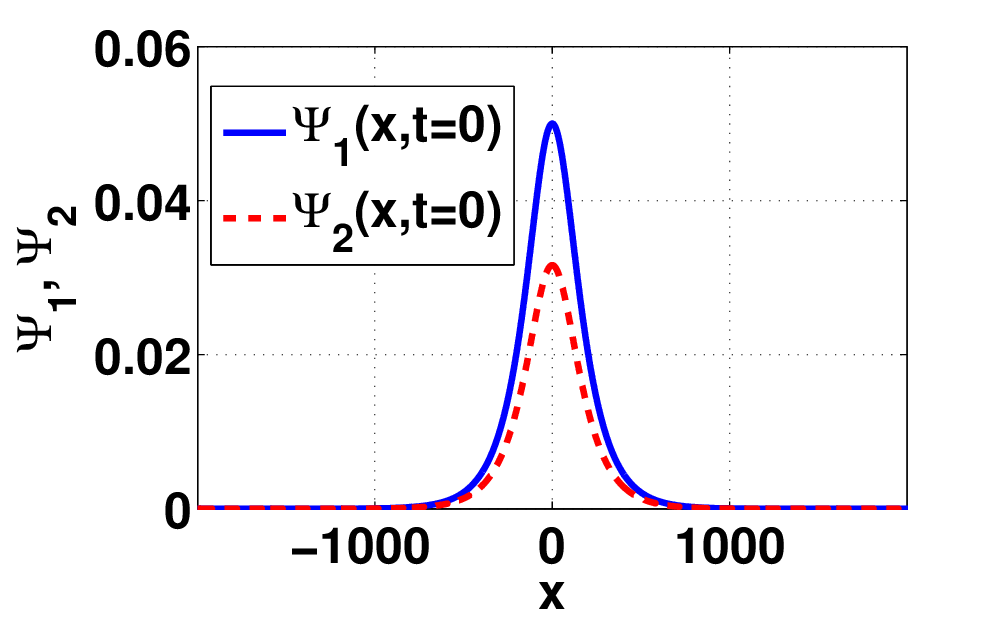}\\
\includegraphics[scale=0.4]{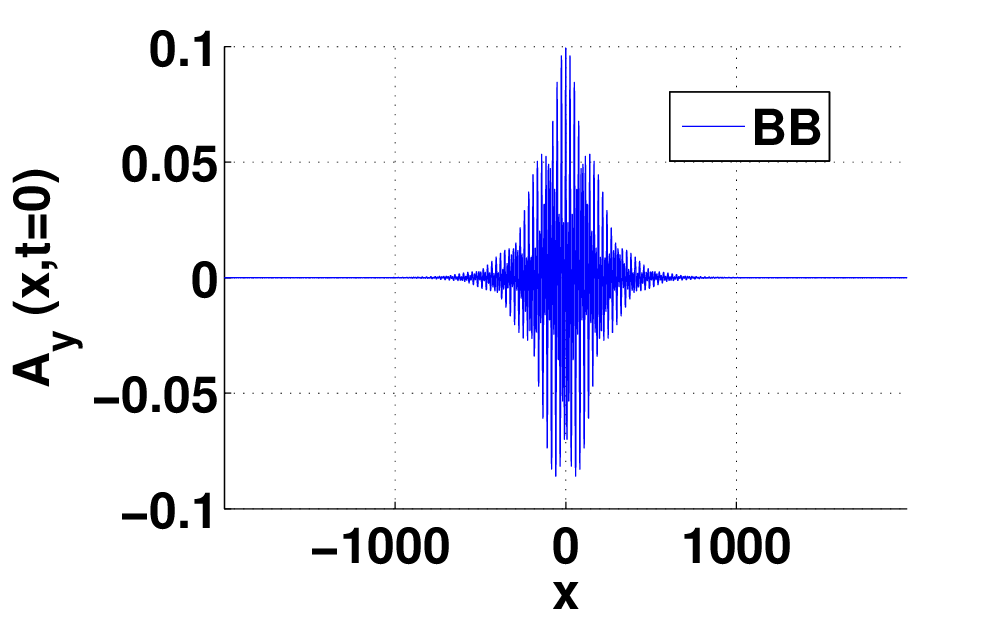}
\end{tabular}
\caption{Case 1: The profiles (at t=0) of the  bright solitons (top panel) and 
dark-bright soliton (bottom panel) for $\Omega=0.8$ . The parameters used are $v_g= 0.4572$,
$\omega_1=0.1454$ ($k_1=0.4932$) and $\omega_2= 0.77414$ ($k_2=0.2621$). 
Also, we use $\epsilon=0.1$ and $\Psi_{1,0}=0.05$.}
\label{fig:BB_sol}
\end{figure}
In Fig.~\ref{fig:BB_sol} we show the profiles (at t=0) of the bright (LCP mode)
and bright (RCP mode) solitons in the absence of coupling (top panel) as well as
the bright-bright soliton (bottom panel), for $\Omega=0.8$.
Now, the group velocity of the BB soliton (common for both components) is
$v_g= 0.4572$, which occurs when the angular frequencies for the modes $\Psi_1$
and $\Psi_2$ take, respectively, the values $\omega_1=0.1454$ ($k_1=0.4932$) and
$\omega_2= 0.77414$ ($k_2=0.2621$).
\begin{figure}[tbp]
\begin{tabular}{cc}
\includegraphics[scale=0.4]{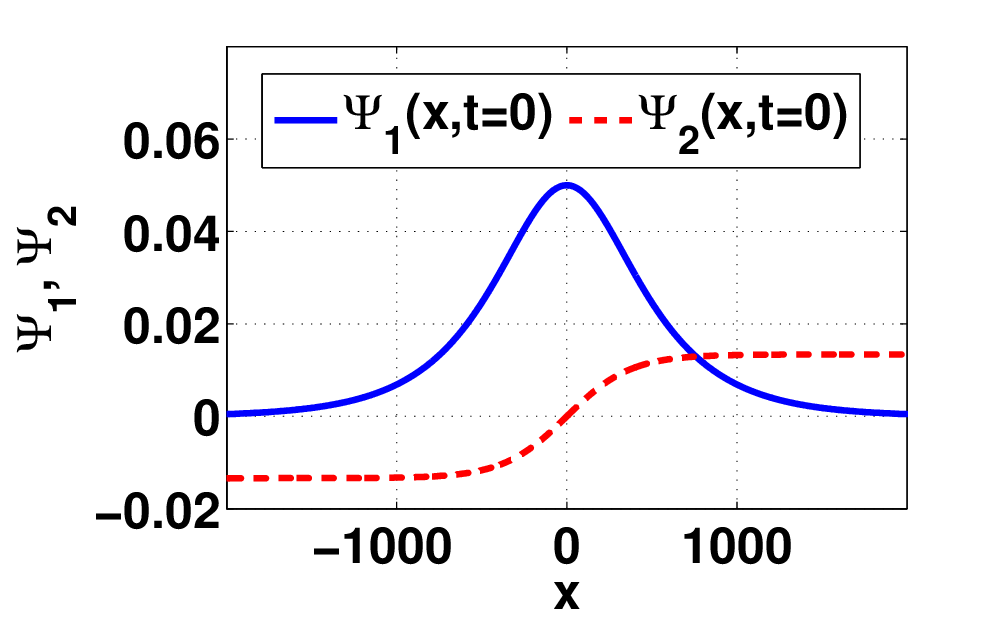}\\
\includegraphics[scale=0.4]{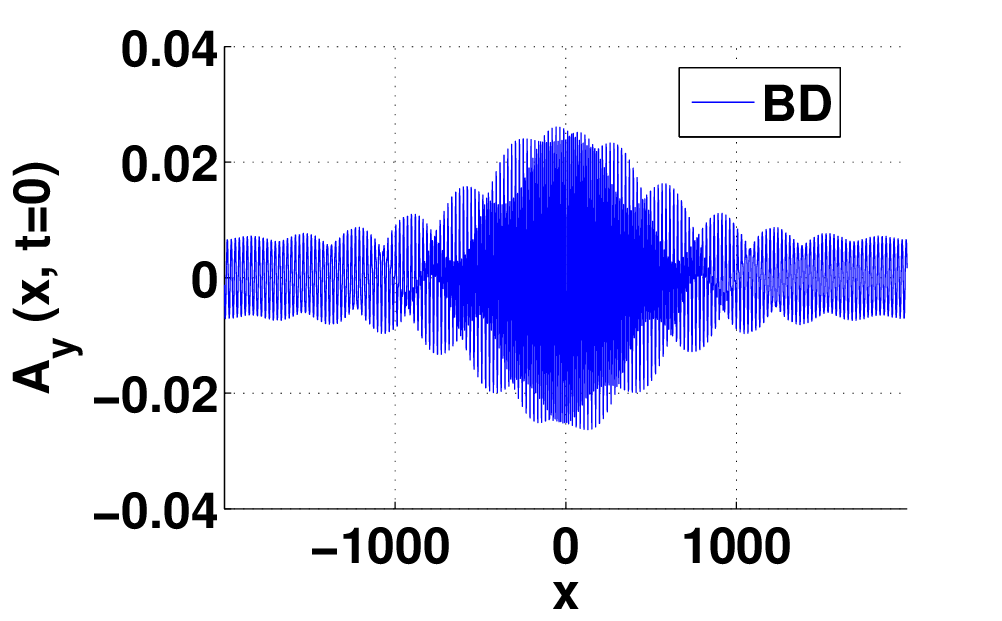}
\end{tabular}
\caption{Case 1: The profiles (at t=0) of the bright and dark solitons (top panel) and bright-dark soliton (bottom panel) for $\Omega=0.8$ . The parameters used are $v_g= 0.4647$,
$\omega_1=0.1743$ ($k_1=0.5558$) and $\omega_2= 0.744$ ($k_2=0.2677$). Also, we use $\epsilon=0.1$ and $\Psi_{1,0}=0.05$.}
\label{fig:BD_sol}
\end{figure}

\textbf{3}.~ When $\alpha_1>0$ and $\alpha_2<0$  then we have a case of coupled solitons, a  bright soliton and a dark soliton, whose exact analytical form are given by Eqs.~(\ref{eq:bs2})-(\ref{eq:ds2}),
where the soliton amplitude parameters $\eta_{1,2}$ and the inverse width $b$ are connected via the following equations:
\begin{eqnarray}
&&\eta_1=(\alpha_2-\alpha_1 \lambda_1)\Psi_{1,0}^2,
\label{eq:eta12}\\
&&\eta_2=-\alpha_1\Psi_{1,0}^2,
\label{eq:eta22}\\
&&b^2=-\alpha_2 \Psi_{1,0}^2.
\label{eq:bb2}
\end{eqnarray}
Now, we can  approximate the vector potential $A_y(x,t)$, in terms of the
original coordinates $x$ and $t$, as in the Eq.~(\ref{eq:darkbright1}) where, in
this case,
\begin{eqnarray}
\!\!\!\!\!\!\!\!
&&\Psi_1=\Psi_{1,0}\sech[\epsilon b(x-v_g t)],
\label{eq:dbsol12}\\
\!\!\!\!\!\!\!\!
&&\Psi_2=\sqrt{\left|\frac{g_{11}}{g_{22}}\right|\alpha_1}\Psi_{1,0}\tanh[\epsilon b(x-v_g t)].
\label{eq:dbsol22}
\end{eqnarray}
In this case, the solution amplitude $A_0$ and the frequencies $\omega_{0j}$ ($j=1,2$) 
are given by Eqs.~(\ref{Ay01})-(\ref{f01}).
%
\begin{figure*}[tbp]
\begin{tabular}{ccc}
&\includegraphics[scale=0.45]{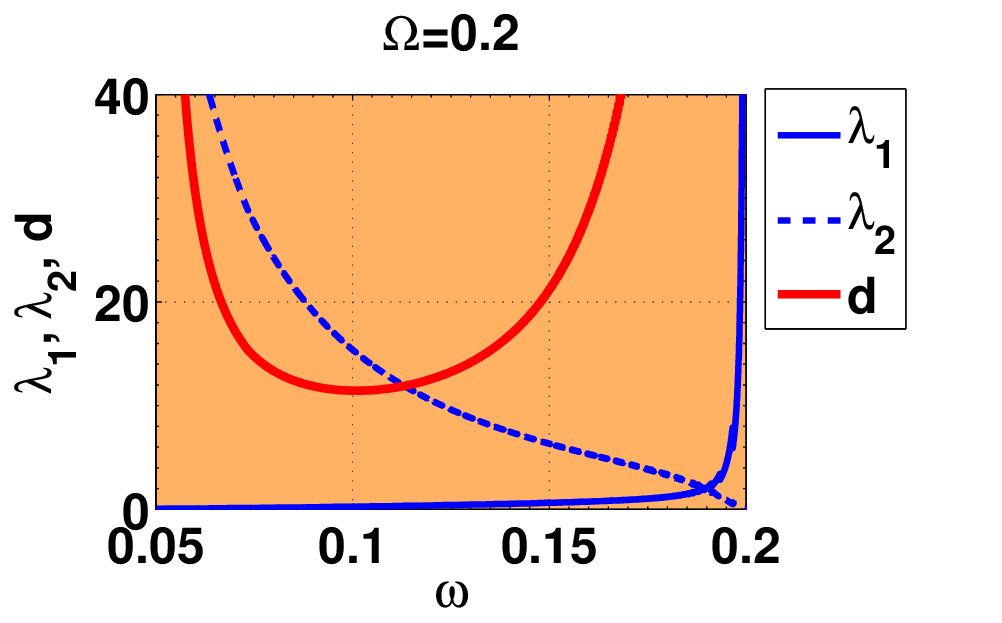}&\includegraphics[scale=0.45]{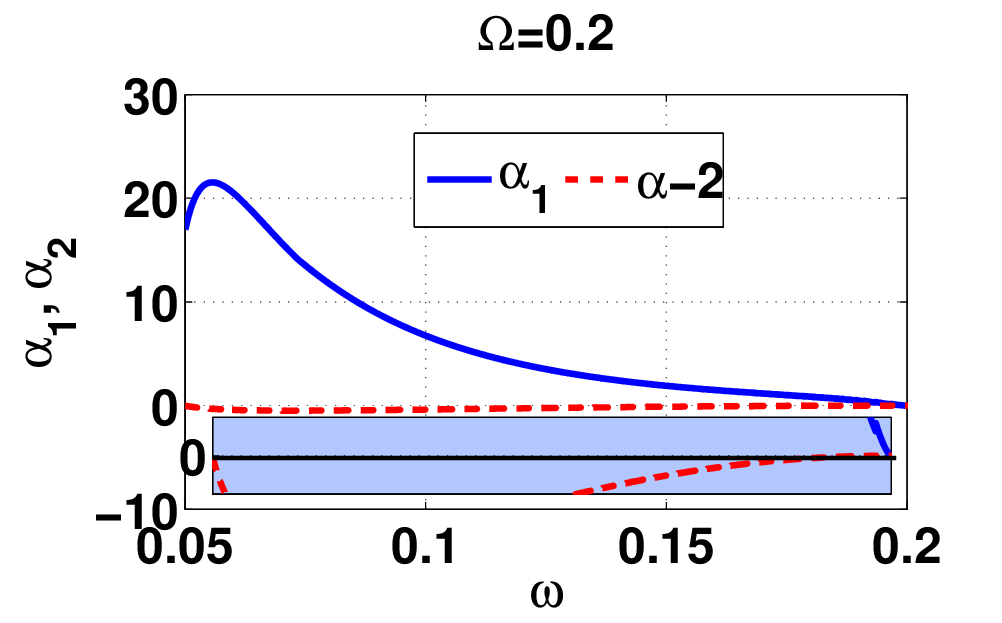}\\
&\includegraphics[scale=0.45]{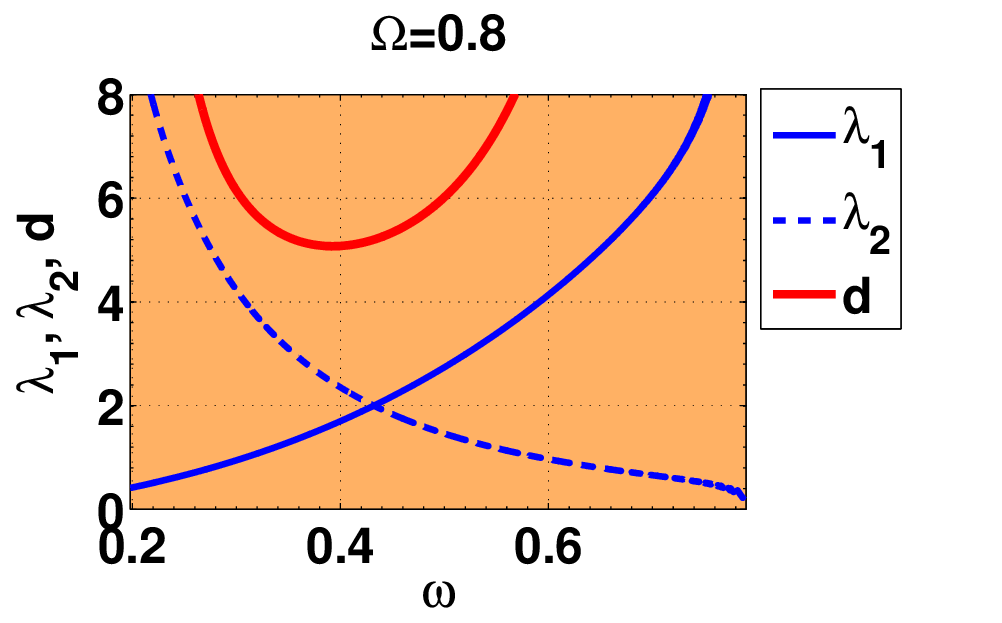}&\includegraphics[scale=0.45]{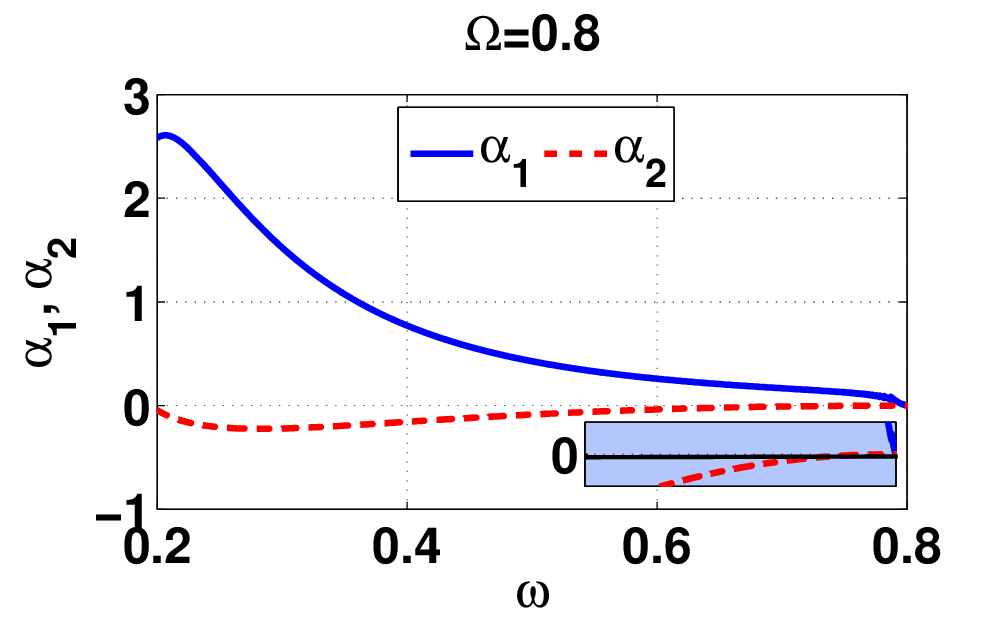}\\
&\includegraphics[scale=0.45]{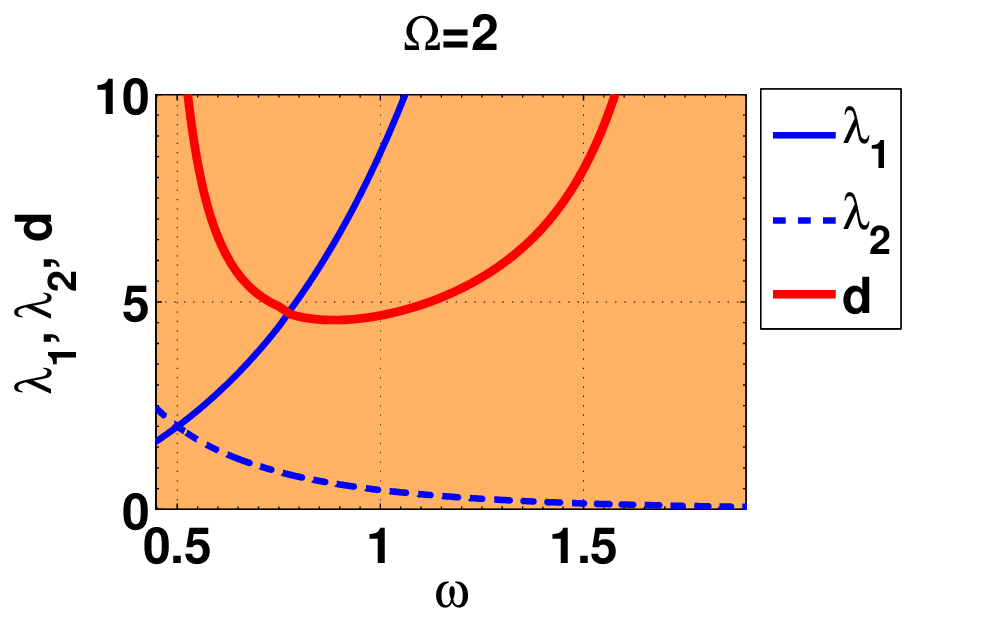}&\includegraphics[scale=0.45]{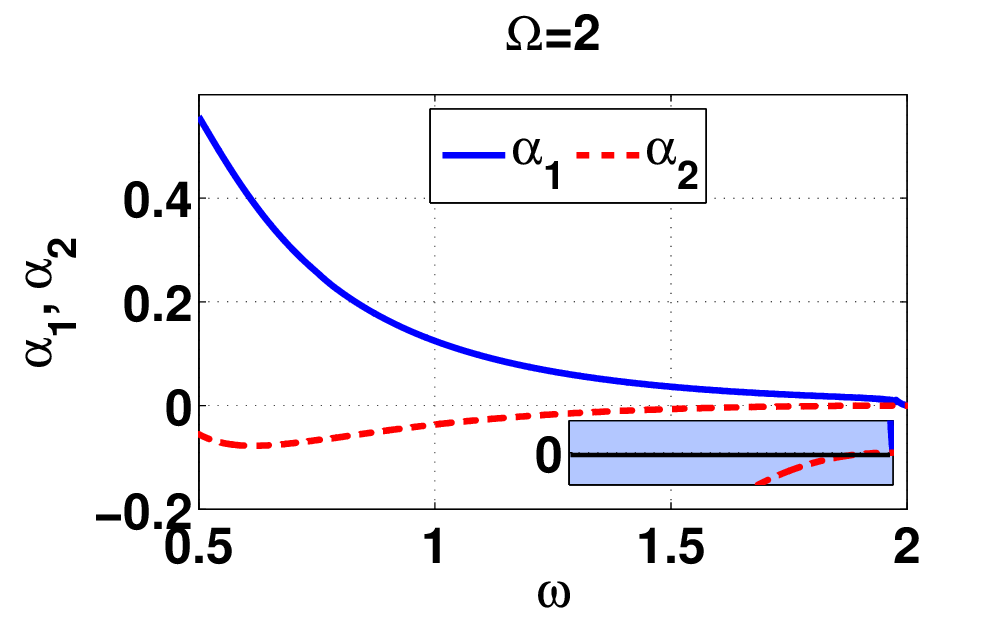}\\
\end{tabular}
\caption{Same as Fig.~\ref{fig:INTADcoeff}, but for Case 2.}
\label{fig:INTCDcoeff}
\end{figure*}
\begin{figure*}[tbp]
\begin{tabular}{ccc}
&\includegraphics[scale=0.45]{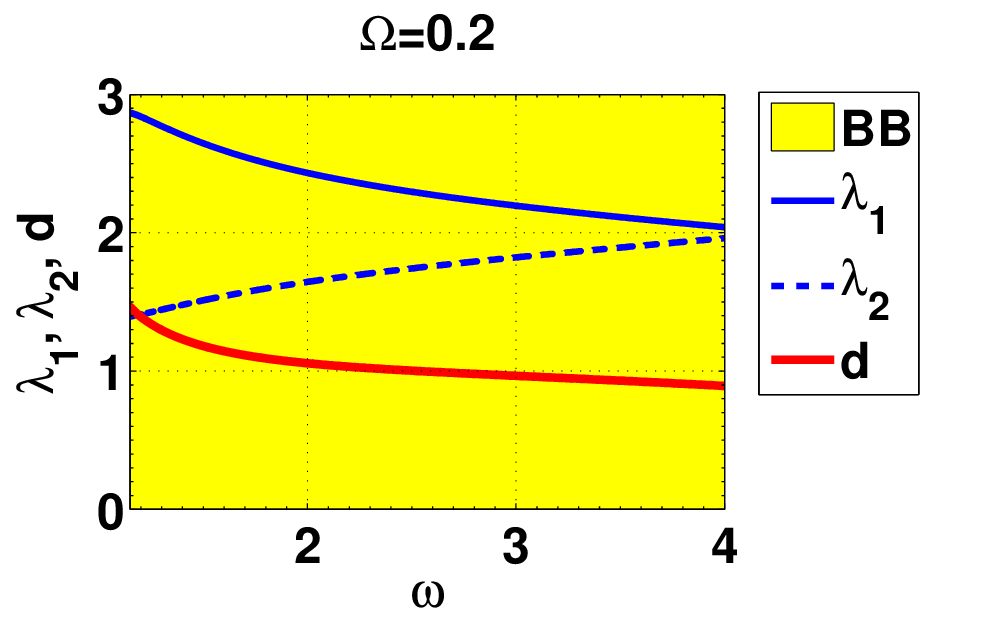}&\includegraphics[scale=0.45]{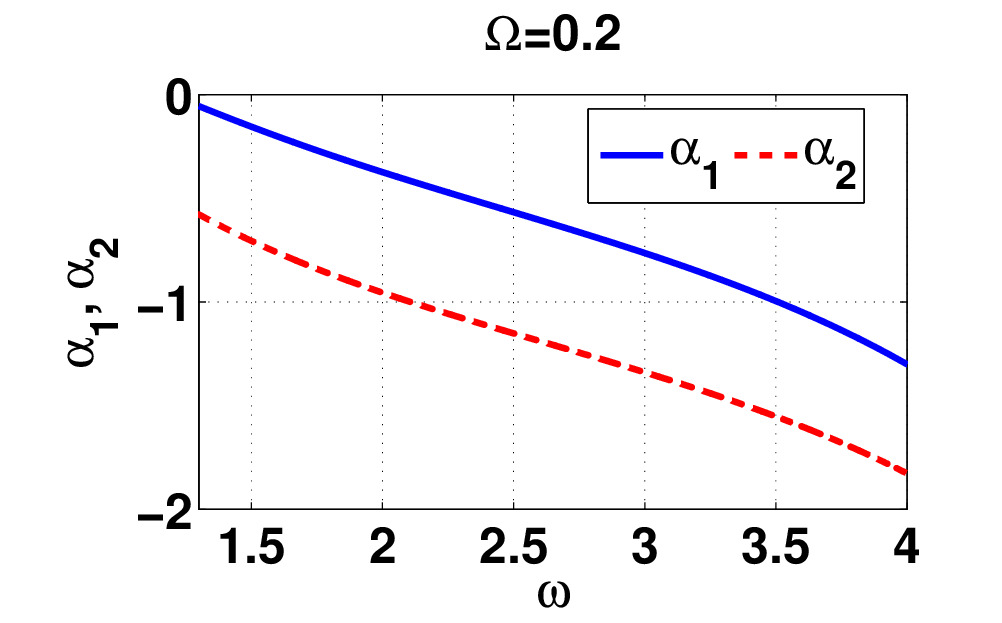}\\
&\includegraphics[scale=0.45]{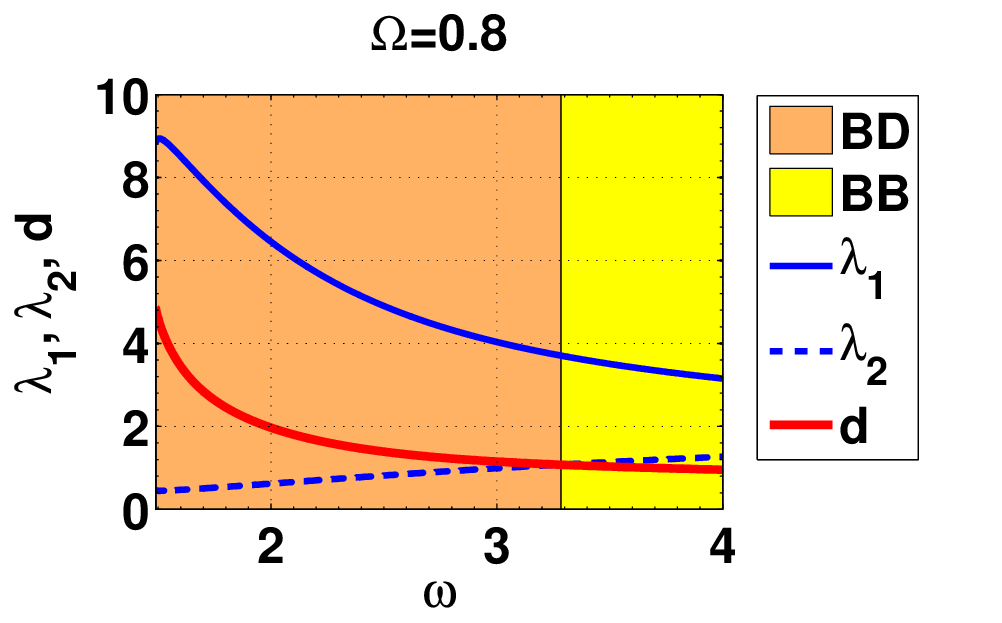}&\includegraphics[scale=0.45]{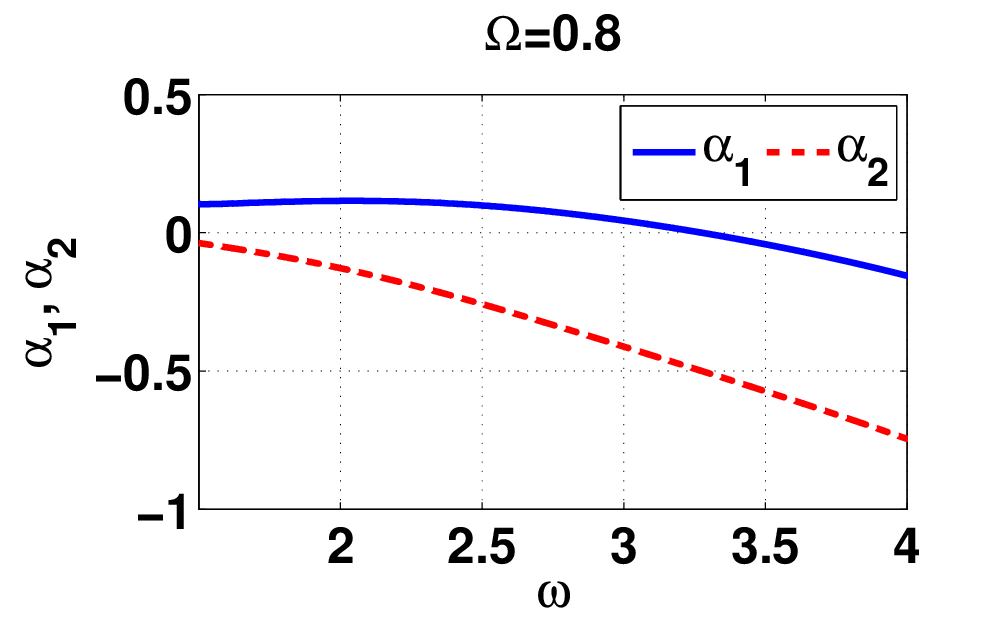}\\
&\includegraphics[scale=0.45]{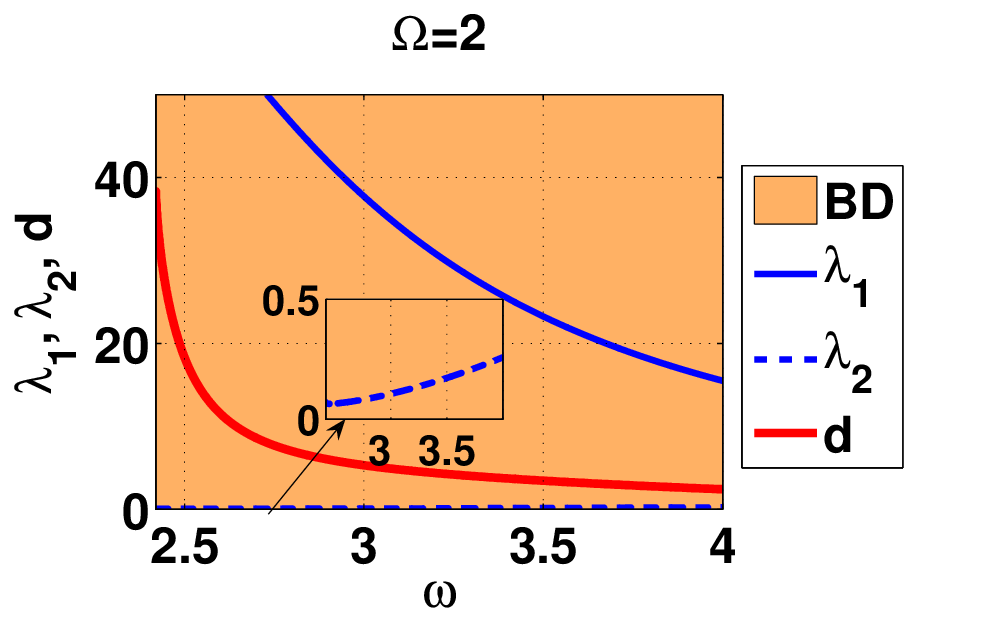}&\includegraphics[scale=0.45]{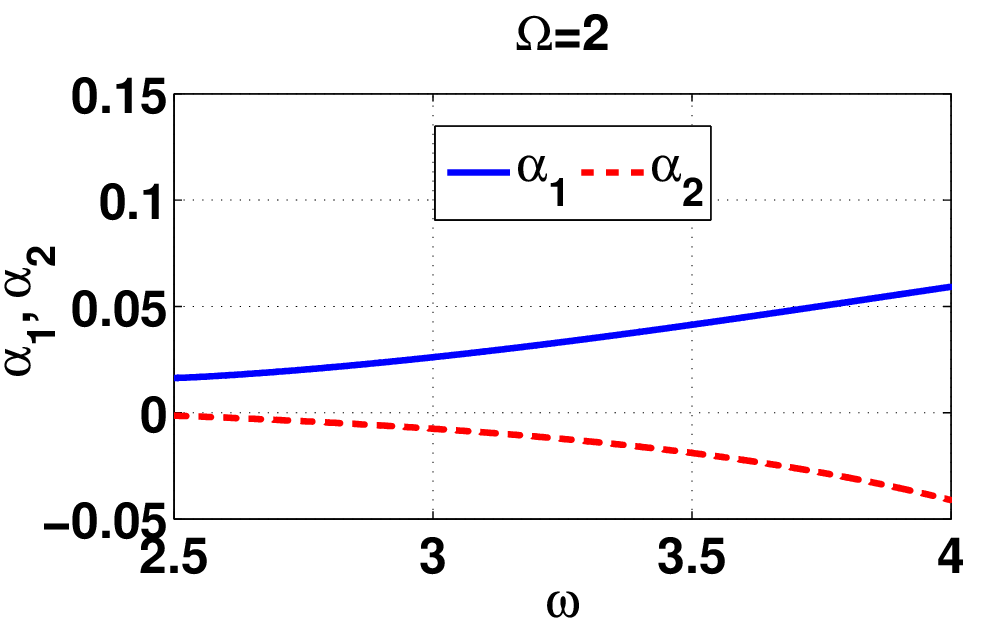}\\
\end{tabular}
\caption{Same as Fig.~\ref{fig:INTADcoeff}, but for Case 3.
}
\label{fig:INTCRcoeff}
\end{figure*}
%

In Fig.~\ref{fig:BD_sol} we show the profiles (at t=0) of the bright (LCP mode) 
and dark (RCP mode) solitons  in the absence of coupling (top panel) as well 
as the bright-dark soliton (bottom panel), for $\Omega=0.8$.
Now, the group velocity of the BD soliton (common for both components) is  
$v_g= 0.4647$, which occurs when the angular frequencies for the modes
$\Psi_1$ and $\Psi_2$ take, respectively, the values $\omega_1=0.1743$ 
($k_1=0.5558$) and $\omega_2= 0.744$ ($k_2=0.2677$).
%
\begin{center}
\label{test}
\underline{\textbf{TABLE 1}}\\
\begin{tabular}{|c|c|c |}
  \multicolumn{3}{c}{~~~~$\alpha_1$}  \\
\cline{2-3}
 \multicolumn {1}{c} {~$\alpha_2$} ~ \vline &\textbf{ -}& \textbf{+} \\
\hline
\textbf{ -}&~~~~\textbf{BB}~~~ &~~~ \textbf{BD}~~~ \\
\hline
 \textbf{+}&  & ~~~\textbf{DB} ~~~\\
\hline
\end{tabular}
\\
\end{center}
The above table shows the existence of vector solitons according to the sign 
of the coefficients $\alpha_1$ and $\alpha_2$, for the case 1. Moreover, as 
are depicted in Fig.~\ref{fig:INTADcoeff}, for $\Omega=0.8$, the above solutions 
are defined only for a narrow frequency band $(0-0.129)$, $(0.136-0.166)$ and  
$(0.17-0.198)$ for dark-bright (DB), bright-bright (BB) and bright-dark (BD) 
interactions , respectively.
Also, as is observed in the Fig.~\ref{fig:INTADcoeff}, as $\Omega$ is decreased 
(increased) lead to an increase (decrease) of the width of the DB band where 
DB solitons can be formed, while the width of BB and BD band is decreased (increased).
As seen in this figure, for $\Omega=0.2$ the DB, BB and BD band possess, approximately, 
the  $80\%$, $ 10\%$ and $ 6\%$ of I-band, respectively. Notice that for $\Omega=0.8$ 
($\Omega=2$) the  DB, BB and BD band possess, the  $ 65\%$ ($47\%$), $15\%$ ($ 21\%$) 
and $14\%$ ($30\%$) of I-band, respectively.

\subsection{Case 2: Bright-dark solitons in bands II and RCP.}
In this case $s=-1$, we study the coupling between a  propagating soliton, 
with a frequency that is in band  II, and a propagating soliton, with a 
frequency that is in RCP band. In Fig.~\ref{fig:INTCDcoeff} we depict the 
dependence of the parameters $\lambda_1$, $\lambda_2$ and $d$, as a function 
of the normalized frequency $\omega$ (for  $\Omega=0.2$,  $\Omega=0.8$ and $\Omega=2$).
%
Now, the coefficients $\alpha_1$ and $\alpha_2$ take the following form:
\begin{eqnarray}
&&\alpha_1=\frac{\lambda_2+d}{1+d\lambda_1},
\label{eq:mu}\\
&&\alpha_2=-\frac{\lambda_1 \lambda_2-1}{2(1+d\lambda_1)}.
\label{eq:nu}
\end{eqnarray}
Following our previous considerations, here we have a case of coupled bright-dark 
solitons where the exact analytical form is given by Eqs.~(\ref{eq:bs2})-(\ref{eq:ds2}).
The soliton amplitude parameters $\eta_{1,2}$ and the inverse width $b$ are connected 
via the following equations:
\begin{eqnarray}
&&\eta_1=-(\alpha_2+\alpha_1 \lambda_1)\Psi_{1,0}^2,
\label{eq:eta122}\\
&&\eta_2=-\alpha_1\Psi_{1,0}^2,
\label{eq:eta222}\\
&&b^2=-\alpha_2 \Psi_{1,0}^2.
\label{eq:bb22}
\end{eqnarray}
The analytical form for vector potential $A_y(x,t)$ and the solitons parameters
are same with the previous case, namely, in the interaction in bands I and RCP
[c.f Eqs.~(\ref{eq:dbsol12})-(\ref{eq:dbsol22})].
\begin{figure}[tbp]
\includegraphics[scale=0.45]{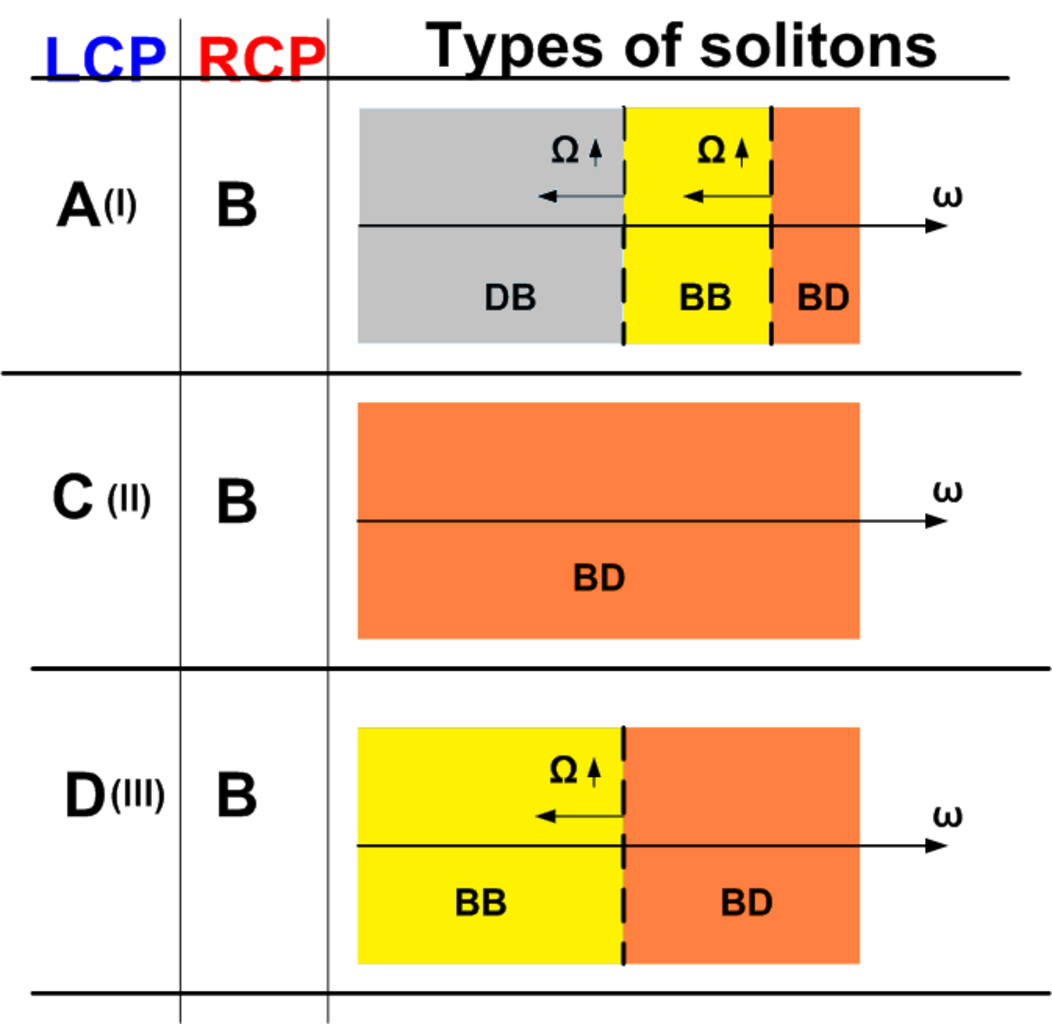}\\
\caption{Different types of solitons as a function of the angular (normalized) frequency  
$\omega$ are depicted in the table, for the different values of
$\Omega$.}
\label{fig:TofS}
\end{figure}
\subsection{Case 3: Solitons in bands III and RCP.}
Finally, we study the coupling between a propagating soliton, with a frequency that is in band III, and a   propagating soliton, with a frequency that is in RCP band. Now, $s=+1$ (cf. Fig.~\ref{fig:gvel}) while the other dispersion and nonlinearity coefficients are shown in Fig.~\ref{fig:INTCRcoeff} as functions of the normalized frequency $\omega$ for $\Omega=0.2$,  $\Omega=0.8$ and $\Omega=2$.
Also, as is observed in the Fig.~\ref{fig:INTCRcoeff}, as $\Omega$ is increased 
lead to an increase  of the width of the BD band where BD solitons can be formed, 
while the width of BB is decreased.
\subsection{Discussions}
The analysis of previous section end in some important results which confirm our expectations. First of all in the Case 1 (Solitons in bands I and RCP) we obtain three types of vector solitons, namely bright-bright, bright-dark and dark-bright. Also, we observe that for low frequencies and $\Omega$ values the width of the dark-bright band where dark-bright solitons can be formed possesses the larger percentage of the band. However as $\Omega$ is increased the occurrence of the dark-bright type is decreased while the width of the bright-bright and mainly also that of the bright-dark category are increased. In Case 2 (Case 2: Bright-dark solitons in bands II and RCP) for all the frequencies and $\Omega$ values, we have only the existence of bright-dark solitons. Finally, in Case 3 (Solitons in bands III and RCP) we have only bright-bright solitons in all the frequencies for low $\Omega$ values. As $\Omega$ is increased bright-dark solitons appeared which dominate in all frequency band for the large values of $\Omega$.

According to the table in Fig.~\ref{fig:TofS}, we found that bright-dark solitons 
exist in all frequency bands for all values of $\Omega$. Furthermore, we observe 
bright-bright solitons in bands (I-RCP) and (III-RCP) for the small values of $\Omega$. 
Also, we found that the dark-bright solitons exist only in I-RCP frequency band for 
small values of $\omega$ and $\Omega$.

Generally our detailed analysis can support various coupled NLS systems in 
magnetized plasmas with the equal group velocities between the propagating waves. 
Our analytical investigation reveals the existence of various type vector solitons 
taking into account the certain frequency bands and the strength of the magnetic 
field. It is very important to note that from our suggested method the capability 
control of a solitonic mode e.g. LCP (RCP) from the change the amplitude or width 
of other mode RCP (LCP) emerges.

Although our present work has an analogous mathematical settings with the works of Refs. \cite{ IKGVrogue}, \cite{veldes}, \cite{NL1},  and \cite{NL3}  it has serious differences as:
a) The physical background in this paper is different compared to Refs.  \cite{veldes}, \cite{NL1},  and \cite{NL3}  . In Refs.  \cite{NL1} and \cite{NL3}  the study was carried out in the framework of an unmagnetized plasma while in the Ref. \cite{veldes} the physical background was completely different where a nonlinear composite right left handed transmission line was studied, b) With regards to Ref. \cite{ IKGVrogue}, where the physical background is same, namely the magnetized plasma, in this paper we study the interaction between left hand circularly polarized (LCP) and right hand circularly polarized (RCP) solitons while in Ref. \cite{ IKGVrogue}, we investigated only the beam-plasma interaction, c) As previously mentioned, in this paper, we investigate the interaction between of a LCP and RCP electromagnetic wave which have the same group velocity in different frequency bands. In Refs \cite{NL1} and \cite{NL3} the electrostatic wave packets are studied in a regime with different group velocities, d)	In this work, we study the vector soliton solutions in different frequency bands and for different magnetic fields. The analysis in Ref. \cite{veldes} was carried out only for different frequency bands, in Ref. \cite{NL1}, the instability growth rate was investigated while in the Ref. \cite{NL3}, the analysis was focused on the values of the $\kappa$-index of k-distribution function.

\section{Conclusion }
In conclusion, we have performed analytical techniques to study the
existence of coupled left- and right-circularly polarized  solitons in magnetized plasmas.
Our analysis started with the derivation of a closed system of scalar 
equations which the govern of the propagation of LCP and RCP EM waves in magnetized plasmas. 
Upon the linear limit, the dispersion relation has been derived and we have shown that 
there are  LCP- and RCP modes, with the same group velocity, which can be propagate  
for specific frequency bands.
Next, in the nonlinear regime we used the multiple scale perturbation method in order to 
investigated the coupling between LCP- and RCP modes.
In this framework, we obtain,
a system of two coupled nonlinear Schr\"{o}dinger (NLS) equations for the unknown 
vector potential envelope functions. The above system of equations was used in order 
to predict the existence  of coupled left- and right-circularly polarized  solitons 
in magnetized plasmas, of the bright-bright, bright-dark and dark-bright type.

Thus, we found vector solitons of the same type (namely, bright-bright) as well as ones of the mixed type (namely, bright-dark and dark-bright) in specific frequency bands. Also, the values of $\Omega$ change the width of frequency band where the vector solitons can be formed. Thus, we found vector solitons of the type dark-bright, bright-bright and bright-dark in the bands I and RCP where the increase of the $\Omega$ lead to a decrease of the width of the band of dark-bright solitons with simultaneous increase of the width of the band of the bright-bright and bright-dark solitons. In the bands II and RCP only bright-dark solitons were found. Finally, we found bright-dark and bright-bright solitons in the bands III and RCP where the increase of the $\Omega$ lead to a decrease of the width of the band of bright-bright solitons with simultaneous increase of the width of the band of the bright-dark solitons.

Our results should be of interest to the plasma science community, where
beam-plasma interactions are a hot topic of research nowadays, both
theoretically and in the laboratory (laser plasma interaction experiments),
and also in nonlinear optics, where coupled beam propagation is an ubiquitous
point of focus. Beyond these physical settings, coupled NLS equations such as
our Eqs. (24)-(25) arise in various contexts, ranging from hydrodynamics to
supraconductivity (Bose-Einstein condensates) and left-hand materials, to
mention a few. The mathematical setting and the methodology of our study is thus
expected to apply in a wider context.

\bigskip

\section*{Acknowledgements}
Author IK and NL gratefully acknowledges financial support from Khalifa
University (United Arab Emirates) via the
project CIRA-2021-064 (8474000412).
IK also acknowledges support from KU via the project  FSU-2021-012 (8474000352),
as well as from KU Space and Planetary Science Center (KU SPSC),
Abu Dhabi, United Arab Emirates, via grant No. KU-SPSC-8474000336.

This work was completed during a research visit by author IK
to the Physics Department, National and Kapodistrian University of Athens, Greece.
During the same period, author IK also held an Adjunct Researcher status at the
Hellenic Space Center, Greece. The hospitality of both hosts, represented by
Professor D.J. Frantzeskakis and (HSC Director) Professor I. Daglis,
respectively, is warmly acknowledged.

Author GV acknowledges support from ADEK
(Abu Dhabi Education and Knowledge Council, currently ASPIRE-UAE) AARE
(Abu Dhabi Award of Research Excellence) grant (AARE18-179) during a research
visit in 2022.
Hospitality from the host (Khalifa University) during that visit is gratefully
acknowledged.


\section{Declarations}

\subsection{Funding}

Authors IK and NL gratefully acknowledge financial support from
Khalifa University of Science and Technology, Abu Dhabi, United Arab Emirates,
via the project CIRA-2021-064 (8474000412) (PI Ioannis Kourakis).
IK also acknowledges financial support from KU via the project FSU-2021-012
(8474000352) (PI Ioannis Kourakis) as well as from KU Space and Planetary
Science Center, via grant No. KU-SPSC-8474000336
(PI Mohamed Ramy Mohamed Elmaarry).

\subsection{Competing Interests}

The authors have no relevant financial or non-financial interests to disclose.

\subsection{Author Contributions}

G. P. Veldes carried out the algebraic work, contributed to the methodology,
  software development, and numerical analysis.
N. Lazarides contributed to the methodology and reviewed the manuscript.
D. J. Frantzeskakis contributed to the concept and design, and reviewed the
  manuscript.
I. Kourakis contributed to the problem conception, project design, launch and
  methodology, reviewed the manuscript, and managed project funding.
All authors read and approved the final manuscript.

\subsection{Data Availability}

The datasets generated during and/or analysed during the current study are
available from the corresponding author on reasonable request.

\end{document}